\documentclass[usegraphicx,usedcolumn,usenatbib]{mn2e}
\usepackage{astbibref}
\usepackage{color,epsfig,amssymb,longtable}
\usepackage{array,colortbl,lscape}

\begin{document}
\title[Abundances of circumnuclear star forming regions]
{The metal abundance of circumnuclear star forming regions in early type spirals. Spectrophotometric observations.}
\author[A.I. D\'\i az et al.]{\'Angeles~I.~D\'{\i}az$^{1}$\thanks{On sabbatical leave at IoA, Cambridge; angeles.diaz@uam.es},
Elena Terlevich$^{2}$\thanks{Research Affiliate at IoA}, 
Marcelo Castellanos$^1$\thanks{At present at the Intituto de Estructura de la Materia, CSIC, Spain} and 
Guillermo H\"agele$ ^{1}$\thanks{PhD fellow of Ministerio de Educaci\'on y Ciencia, Spain}\\
        $^1$Departamento de F\'\i sica Te\'orica, C-XI, Universidad Aut\'onoma de Madrid, 28049 Madrid, Spain\\
        $^2$INAOE, Tonantzintla, Apdo. Postal 51, 72000 Puebla, M\'exico\\
}
\date{Accepted 
      Received ;
      in original form June 2007 }

\pagerange{\pageref{firstpage}--\pageref{lastpage}}
\pubyear{2007}

%\begin{document}

\maketitle

\label{firstpage}

\begin{abstract}

  We have obtained long-slit observations in the optical and near infrared of 12 circumnuclear HII regions (CNSFR) in the early type spiral galaxies NGC~2903, NGC~3351 and NGC~3504 with the aim of deriving their chemical abundances.
Only for one of the regions, the [SIII] $\lambda$ 6312  \AA\ was detected providing, together with the nebular [SIII] lines at $\lambda\lambda$ 9069, 9532 \AA , a value of the electron temperature of T$_e$([SIII])= 8400$^{+ 4650}_{-1250}$K. A semi-empirical method for the derivation of abundances in the high metallicity regime is presented. 

We obtain abundances which are comparable to those found in high metallicity disc HII regions from direct measurements of electron temperatures and consistent with solar values within the errors. The region with the highest oxygen abundance is R3+R4 in NGC~3504, 12+log(O/H) = 8.85, about 1.5 solar if the solar oxygen abundance is set at the value derived by Asplund et al. (2005), 12+log(O/H)$_{\odot}$ = 8.66$\pm$0.05. Region R7 in NGC~3351 has the lowest oxygen abundance of the sample, about 0.6 times solar. In all the observed CNSFR the O/H abundance is dominated by the O$^+$/H$^+$ contribution, as is also the case for high metallicity disc HII regions. For our observed regions, however, also the  S$^+$/S$^{2+}$ ratio is larger than one, contrary to what is found in high metallicity disc HII regions for which, in general, the sulphur abundances are dominated by S$^{2+}$/H$^+$. 

The derived N/O ratios are in average larger than those found in high metallicity disc HII regions and they do not seem to follow the trend of N/O vs O/H which marks the secondary behaviour of nitrogen. On the other hand, the S/O ratios span a very narrow range between 0.6 and 0.8 of the solar value.
 
As compared to high metallicity disc HII regions, CNSFR show values of the O$_{23}$ and the N2 parameters whose distributions are shifted to lower and higher values respectively, hence, even though their derived oxygen and sulphur abundances are similar, higher values would in principle be obtained for the CNSFR  if pure empirical methods were used to estimate abundances. CNSFR also show lower ionisation parameters than their disc counterparts, as derived from the [SII]/[SIII]. Their ionisation structure also seems to be different with CNSFR showing radiation field properties more similar to HII galaxies than to disc high metallicity HII regions. 
%\ojo\ lo que sigue, Roberto sugiere sacarlo del abstract \ojo\ The possible contamination of their spectra from hidden low luminosity AGN and/or shocks, as well as the probable presence  of more than one velocity component in the ionised gas corresponding to kinematically distinct systems, sould be further investigated. 

. 

\end{abstract}

\begin{keywords}
galaxies:  abundances -- galaxies: HII regions, abundances
\end{keywords}

\section{Introduction}

The inner ($\sim$ 1Kpc) parts of some spiral galaxies show high star formation rates 
and this star formation is frequently arranged in a ring or pseudo-ring pattern around their nuclei. This fact seems to correlate with the presence of bars and, in fact, computer models which simulate the behaviour of gas in galactic potentials have shown that nuclear rings may appear as a consequence of matter infall owing to resonances present at the bar edges \citep{1985A&A...150..327C,1992MNRAS.259..328A}.

In general, Circumnuclear Star Forming Regions (CNSFR), also referred to as ``hotspots'', and luminous and large disc HII regions are very much alike, but look more compact and show higher peak surface brightness  \citep{1989AJ.....97.1022K}.  In many cases they contribute substantially to the emission of the entire nuclear region. 

Their large H$\alpha$ luminosities, typically higher than 10$^{39}$ erg s$^{-1}$, point to relatively massive star clusters as their ionisation source, which minimizes the uncertainties due to small number statistics when applying population synthesis techniques \citep[see e.g.][]{2002A&A...381...51C}. These regions then constitute excellent places to study how star formation proceeds in high metallicity, high density circumnuclear environments.

In many cases, CNSFR show emission line spectra similar to those of disc HII regions. However, they show a higher continuum from background stellar populations as expected from their circumnuclear location, often inside 500 pc from the galaxy centre. In early type spirals CNSFR are also expected to be amongst the highest metallicity regions as corresponds to their position near the galactic bulge. These facts taken together make the analysis of these regions complicated since, in general, their low excitation makes any temperature sensitive line too weak to be measured, particularly against a strong underlying stellar continuum.  In fact, in most cases, the [OIII]$\lambda$ 5007 \AA\ line, which is typically  one hundred times more intense than the auroral [OIII]$\lambda$ 4363 \AA\ one, can be barely seen. 

Yet, despite its difficulty, the importance of an accurate determination of the abundances of high metallicity HII regions cannot be overestimated since they constitute most of the HII regions in early spiral galaxies (Sa to Sbc) and the inner regions of most late  type ones (Sc to Sd) \citep{1989epg..conf..377D,1992MNRAS.259..121V} without  which our description of the metallicity distribution in galaxies cannot be  complete. In particular, the effects of the choice of different calibrations  on the derivation of abundance gradients can be very important since any abundance  profile fit will be strongly biased towards data points at the ends of the  distribution. It should be kept in mind that abundance gradients are widely  used to constrain chemical evolution models, histories of star formation over galactic discs or galaxy formation scenarios. The question of how high is the highest oxygen abundance in the gaseous phase of galaxies is still standing and extrapolation of known radial abundance gradients would point to CNSFR as the most probable sites for these high metallicities. 

Accurate measures of elemental abundances of high metallicity regions are crucial to obtain reliable calibrations of empirical abundance estimators, widely used but poorly constrained, whose choice can severely bias results obtained for quantities of the highest relevance for the study of galactic evolution like the luminosity-metallicity (L-Z) relation for galaxies. CNSFR are also ideal cases to study the behaviour of abundance estimators in the high metallicity regime.

In section 2 of this paper we describe the sample of regions observed and the circumnuclear environment of their host galaxies. Our observations and the data reduction 
procedures are described in section 3. The results are presented in  section 4. The method for the derivation of chemical abundances is described in section 5. Section 6 is devoted to the discussion of our results. Finally, section 7 summarizes the main conclusions of this work.

\section{Sample selection}

We have obtained moderate resolution observations of 12 CNSFR in three ``hot spot" galaxies: NGC~2903, NGC~3351 and NGC~3504 whose main properties are given in Table \ref{propgal}. The three of them are early barred spirals and show a high star formation rate in their nuclear regions. They are quoted in the literature as among the spirals with the highest overall oxygen abundances \citep{1994ApJ...420...87Z,tesisdiego}. 

NGC~2903 is a well studied galaxy. The Paschen $\alpha$ image obtained with the Hubble Space Telescope (HST) reveals the presence of a nuclear ring-like morphology with an apparent diameter of approximately 15 \arcsec = 625 pc \citep{2001MNRAS.322..757A}. This structure is also seen, though less prominent, in the H$\alpha$ observations from \cite{1997A&A...325...81P}. A large number of stellar clusters are identified on  high resolution infrared images in the K' and H bands, which do not coincide spatially with the bright HII regions. A possible interpretation of this is that the stellar clusters are the result of the evolution of giant HII regions \citep[e.g.][]{2001MNRAS.322..757A}. The global star formation rates in the nuclear ring, as derived from  its H$\alpha$ luminosity  is found to be 0.1 \,M$_\odot$\,yr$^{-1}$ by \cite{1997A&A...325...81P} and 0.7 \,M$_\odot$\,yr$^{-1}$ from \citep{2001MNRAS.322..757A}. From CO emission observations, \cite{1997A&A...325...81P} derive a mass of molecular gas (H$_2$) of 1.8\,$\times$\,10$^{8}$\,M$_\odot$ inside a circle 1 Kpc in diameter. 

NGC~3351 is another well known ``hot spot" galaxy  \citep{1967PASP...79..152S}. Early detailed studies of its nuclear regions  \citep{1982A&AS...50..491A} concluded that NGC\,3351  harbours high-mass circumnuclear star formation. In fact, the star formation rate per unit area in the nuclear region is significantly increased over that observed in the disc \citep{1992AJ....103..784D}. \cite{1997AJ....114.1850E} from near infrared photometry in  the J and K bands derive a circumnuclear star formation rate of 0.38\,M$_\odot$\,yr$^{-1}$.  \cite{1997A&A...325...81P}, from the H$\alpha$ emission,  derive a total star formation rate for the circumnuclear region of 0.24\,M$_\odot$\,yr$^{-1}$ and a mass of molecular gas of 3.5\,$\times$\,10$^{8}$\,M$_\odot$ inside a circle of 1.4 Kpc in diameter, from CO emission observations. A recent kinematical study of the CNSFR has been presented in \cite{2007MNRAS.378..163H}.

NGC~3504 is the brightest galaxy in the optically selected catalogue of starburst galactic nuclei in \cite{1983ApJ...268..602B}.
It forms a pair with NGC~3512. Various studies at different wavelengths confirmed that it harbours a very intense nuclear starburst \citep{1989ApJ...346..126D,1990ApJ...364...77P}. Infrared observations in the J and K bands reveal a ring with five discrete clumps of star formation with colours indicating ages of about 10$^7$ yr \citep{1997AJ....114.1850E}. The H$\alpha$ emission from \cite{1997A&A...325...81P} traces a compact ring structure with a radius of 2\arcsec (200 pc) around the nucleus where four separated HII regions can be identified. From the H$\alpha$ emission, these authors derive a global star formation rate for the circumnuclear region of 0.62 \,M$_\odot$\,yr$^{-1}$, while from their CO observations, a molecular gas mass of 21  \,$\times$\,10$^{8}$\,M$_\odot$ 
inside a circle 2.7 Kpc in diameter is derived.

%%%%%%%%%%%%%%%%%%%%%%%%%%%%%%%%%%%%%%%%%%%%%%%%%%%%%%%%%%%%%%%%%%%%%%%%%%%%%%
%
%
%OBSERVATIONS AND DATA REDUCTION
%
%
%%%%%%%%%%%%%%%%%%%%%%%%%%%%%%%%%%%%%%%%%%%%%%%%%%%%%%%%%%%%%%%%%%%%%%%%%%%%%%

\section[]{Observations and data reduction}

%%%%%%%%%%%%%%%%%%%%%%%%%%%%%%%%%%%%%%%%%%%%%%%%%%%%%%%%%%%%%%%%%%%%%%%%%%%%%%%
%           TABLA 1 ----GALAXY SAMPLE AND MAIN PROPERTIES
%
%%%%%%%%%%%%%%%%%%%%%%%%%%%%%%%%%%%%%%%%%%%%%%%%%%%%%%%%%%%%%%%%%%%%%%%%%%%%%%%

\begin{table}
\begin{center}
\caption{The galaxy sample}
 \begin{tabular}{cccc}
\hline
Property              & NGC 2903  & NGC 3351 & NGC 3504  \\
\hline
R. A. (2000)$^a$  &  09 32 10.1 & 10 43 57.7 & 11 03 11.2 \\
Dec (2000)$^a$   & +21 30 03   & +11 42 14  & +27 58 21 \\ 
Morph. Type        & SBbc          & SBb          &     SABab  \\
Distance (Mpc)$^b$
                          &  8.6             &    10         & 20   \\
pc/ \arcsec\         &  42              &    50         &  100   \\
B$_{T}$ (mag)$^a$         
                          &  9.7             &    10.5      & 11.8  \\
E(B-V)$_{gal}$(mag) $^a$  & 0.031 & 0.028 & 0.027 \\
\hline
\multicolumn{4}{l}{$^a$~\cite{1991trcb.book.....D}}\\
\multicolumn{4}{l}{$^b$~NGC 2903: \cite{1984AAS...56..381B}}\\
\multicolumn{4}{l}{~~NGC 3351: \cite{1997ApJ...477..535G}}\\
\multicolumn{4}{l}{~~NGC 3504: \cite{1992ApJ...395L..79K}} \\
\end{tabular}
\label{propgal}
\end{center}
\end{table}

%%%%%%%%%%%%%%%%%%%%%%%%%%%%%%%%%%%%%%%%%%%%%%%%%%%%%%%%%%%%%%%%%%%%%%%%%%%%%%%

Our spectrophotometric observations were obtained with the 4.2m William Herschel Telescope at the Roque de los Muchachos Observatory, in 2001 January 26, using the ISIS double spectrograph, with the EEV12 and TEK4 detectors in  the blue and red arm respectively. The incoming light was split by the dichroic at $\lambda$7500 \AA\ .  Gratings  R300B in the blue arm and R600R in the red arm were used, covering  3400 {\AA} in  the blue ($\lambda$3650 to $\lambda$7000) and 800 {\AA} in the near IR ($\lambda$8850 to $\lambda$9650) and yielding spectral  dispersions of 1.73 {\AA} pixel$^{-1}$ in the blue arm and  0.79 {\AA} pixel$^{-1}$ in the red arm. With a slit width of 1\farcs05,  spectral resolutions of {$\sim$}2.0 {\AA} and 1.5 {\AA} FWHM in the blue and red arms respectively were attained. This is an optimal configuration which allows the simultaneous observation of a given region in both frames in a single exposure. 

The nominal spatial sampling is 0\farcs4 pixel$^{-1}$ in each frame and the average seeing for this night was {$\sim$}1\farcs2. A journal of the observations is given in Table \ref{journal}. 

%%%%%%%%%%%%%%%%%%%%%%%%%%%%TABLAS%%%%%%%%%%%%%%%%%%%%%%%%%%%%%%%%%%%%%%%%%%%%%
%                                                                             
%             TABLA 2  : JOURNAL OF OBSERVATIONS                              
%                                                                             
%%%%%%%%%%%%%%%%%%%%%%%%%%%%%%%%%%%%%%%%%%%%%%%%%%%%%%%%%%%%%%%%%%%%%%%%%%%%%%%
\begin{table*}
\centering
\caption[]{Journal of Observations}
\begin{tabular} {l c c c c c c}
\hline
 Galaxy & Spectral range & Grating &      Disp.       &    Spatial resolution & PA   & Exposure Time \\
            &     (\AA)           &            & (\AA\,px$^{-1}$)  & (\arcsec\,px$^{-1}$) &  ($ ^{o} $) & (sec)   \\
\hline
NGC~2903 & 3650-7000  & R300B &      1.73       &  0.4    &     105    &   2\,$\times$\,1800 \\
NGC~2903 & 8850-9650  & R600R &      0.79       &  0.4    &     105    &   2\,$\times$\,1800  \\
NGC~2903 & 3650-7000  & R300B &      1.73       &  0.4    &     162    &   2\,$\times$\,1800 \\
NGC~2903 & 8850-9650  & R600R &      0.79       &  0.4    &     162    &   2\,$\times$\,1800 \\
NGC~3351 & 3650-7000  & R300B &      1.73       &  0.4    &      10     &   2\,$\times$\,1800 \\
NGC~3351 & 8850-9650  & R600R &      0.79       &  0.4    &      10     &   2\,$\times$\,1800  \\
NGC~3351 & 3650-7000  & R300B &      1.73       &  0.4    &      38     &   2\,$\times$\,1800 \\
NGC~3351 & 8850-9650  & R600R &      0.79       &  0.4    &      38     &   2\,$\times$\,1800  \\
NGC~3351 & 3650-7000  & R300B &      1.73       &  0.4    &      61     &   2\,$\times$\,1800 \\
NGC~3351 & 8850-9650  & R600R &      0.79       &  0.4    &      61     &   2\,$\times$\,1800  \\
NGC~3504 & 3650-7000  & R300B &      1.73       &  0.4    &     110    &   2\,$\times$\,1800 \\
NGC~3504 & 8850-9650  & R600R &      0.79       &  0.4    &     110    &   2\,$\times$\,1800  \\
\hline
\end{tabular}
\label{journal}
\end{table*}

%%%%%%%%%%%%%%%%%%%%%%%%%%%%%%%%%%%%%%%%%%%%%%%%%%%%%%%%%%%%%%%%%%%%%%%%%%%%%%%

\begin{figure*}
\centering
\includegraphics[width=.41\textwidth,angle=0]{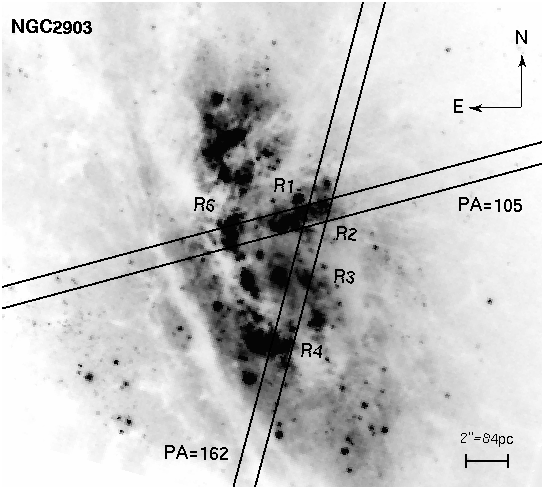}
\hspace{0.2cm}
\includegraphics[width=.41\textwidth,angle=0]{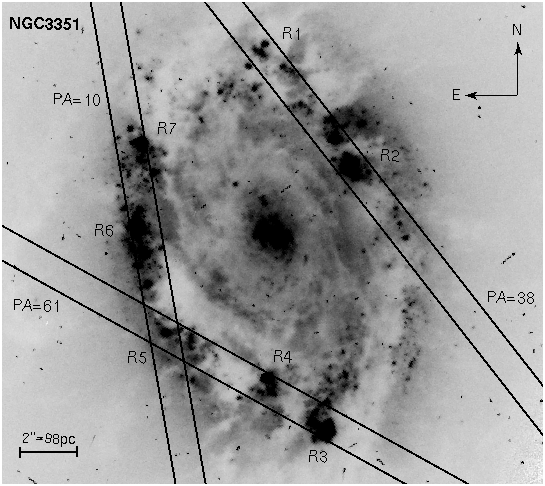}
\hspace{0.2cm}
\includegraphics[width=.41\textwidth,angle=0]{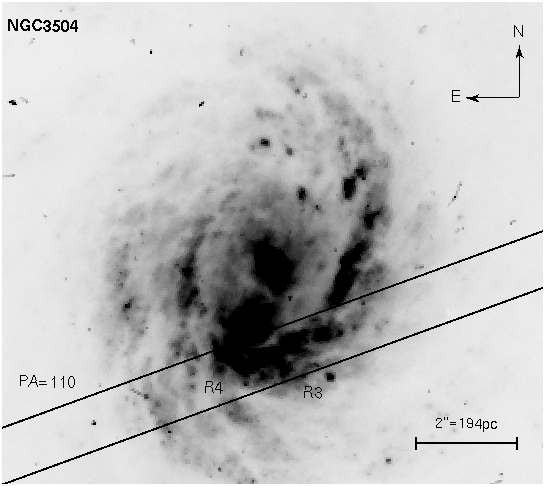}  
\caption[]{Observed CNSFR in each of the galaxies. The different slit positions are superimposed on images taken from the HST archive and obtained with the WFPC2 camera through the F606W filter. The position angles of every slit position are indicated.}
\label{hst-slits}
\end{figure*}

%%%%%%%%%%%%%%%%%%%%%%%%%%%%%%%%%%%%%%%%%%%%%%%%%%%%%%%%%%%%

Two, three and one slit positions in NGC 2903, NGC 3551 and NGC 3504 respectively were chosen to observe a total of 12 CNSFR. Figure \ref{hst-slits} shows the different slit positions superimposed on images obtained with the HST WFPC2 camera and taken from the HST archive.  Some characterstics of the observed regions, as given by  Planesas et al. (1997), from where the identification numbers have also been taken, are listed in Table \ref{prop_cn}.

%%%%%%%%%%%%%%%%%%%%%%%%%%%%%%%%%%%%%%%%%%%%%%%%%%%%%%%%%%%%
%                                                                             
%             TABLA 3  : Characteristics of the CNSFR observed                          
%                                                                             
%%%%%%%%%%%%%%%%%%%%%%%%%%%%%%%%%%%%%%%%%%%%%%%%%%%%%%%%%%%%%%%%%%%%%%%%%%%%%%%

\begin{table*}
\centering
 \caption{Characteristics of the CNSFR observed}
 \begin{tabular}{@{}lcccc@{}}
\hline
\hline
Galaxy & Region & Offsets from centre & Diameter$^a$& F(H$\alpha$)\\
            &            & \arcsec\, \arcsec\       & \arcsec\  &
$\times$10$^{-14}$ erg s$^{-1}$ cm$^{-2}$ \\

\hline
NGC~2903 & R1+R2 & -1.7,+3.5  & 4.0 & 22.8 \\
                 & R3       & -1.6,-0.3   & 2.0 & 2.6  \\
                 & R4       & -0.3,-3.3   & 2.4 & 10.1 \\
                 & R6       & +2.3,+2.4 & 2.4 & 12.7 \\  
NGC~3351 & R1       & +0.4,+6.5 & 2.4 & 12.3 \\
                 & R2       & -2.6,+2.6  & 2.4 & 16.0 \\
                 & R3       & -1.5,-6.5   & 2.4 & 20.7 \\
                 & R4       & +0.5,5.4   & 2.2 & 10.5 \\
                 & R5       & +2.9,-3.5  & 2.4 & 4.9 \\
                 & R6       & +4.1,-0.4  & 2.4 & 6.1 \\
                 & R7       & +4.8,+3.6 & 1.8 & 9.5 \\
NGC~3504 & R3+R4 & -0.5,-1.7   & 1.6 & 21.6 \\

\hline
\multicolumn{5}{l}{$^a$~Size of the circular aperture used to measure fluxes} \\
\end{tabular}
\label{prop_cn}
\end{table*}

%%%%%%%%%%%%%%%%%%%%%%%%%%%%%%%%%%%%%%%%%%%%%%%%%%%%%%%%%

 The data were reduced using the IRAF\footnote{IRAF: the Image Reduction and Analysis Facility is distributed by
  the National Optical Astronomy Observatories, which is operated by the
  Association of Universities for Research in Astronomy, Inc. (AURA) under
  cooperative agreement with the National Science Foundation (NSF).} package following standard methods. The two-dimensional wavelength calibration was accurate to 1 {\AA} in all cases by means of Cu, Ne and Ar calibration lamps. The two-dimensional frames were flux calibrated using four spectroscopic standard stars: Feige~34, BD26+2606, HZ44 and HD84937, observed before and after each programme object with a 3\arcsec\ 
%3\farcs\
width slit. For two of the standard stars: Feige~34 and HZ44, the fluxes have been obtained from the most updated version of the original Oke's spectra \citep{1990AJ.....99.1621O} and cover the 3200 to 9200 {\AA} range. Data between 9200 and 9650 {\AA} have been obtained from stellar atmosphere models. For the other two stars: BD26+2606 and HD84937, the fluxes have been taken from \citet{1983ApJ...266..713O} that cover the whole spectral range. The agreement between the individual calibration curves was better than 5\% in all cases and a weighted mean calibration curve was derived. The spectra were previously corrected for atmospheric extinction using a mean extinction curve applicable to La Palma observing site.

Regarding background subtraction, the high spectral dispersion used in the near infrared allowed the almost complete elimination of the night-sky OH emission lines and, in fact, the observed  $\lambda$9532/$\lambda$9069 ratio is close to the theoretical value of 2.44 in most cases. Telluric absorptions have been removed from the spectra of the regions by dividing by the relatively featureless continuum of a subdwarf star observed in the same night.

%%%%%%%%%%%%%%%%%%%%%%%%%%%%%%%%%%%%%%%%%%%%%%%%%%%%%%%%%%%%%%%%%%%%%%%%%%%%%%%%
%                                                                              
%           Figura 3: Perfiles en Halfa                                        
%                                                                              
%%%%%%%%%%%%%%%%%%%%%%%%%%%%%%%%%%%%%%%%%%%%%%%%%%%%%%%%%%%%%%%%%%%%%%%%%%%%%%%%

\begin{figure*}[!ht]
%\centering
\includegraphics[width=.41\textwidth,angle=0]{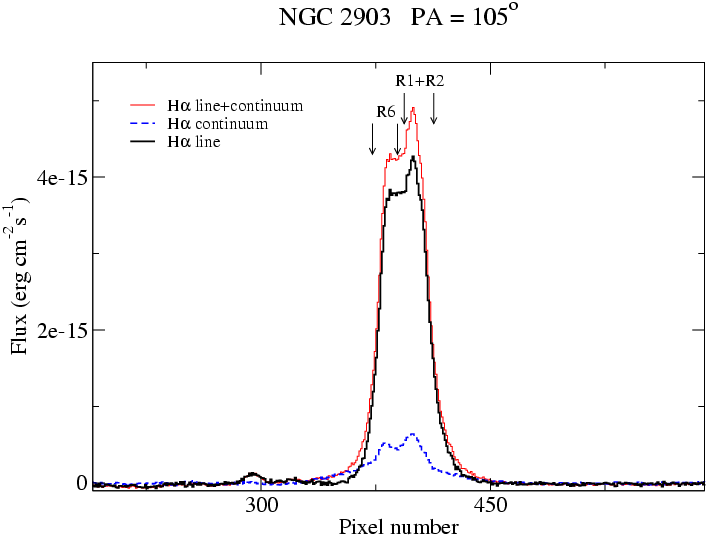}
\hspace{0.2cm}
\includegraphics[width=.41\textwidth,angle=0]{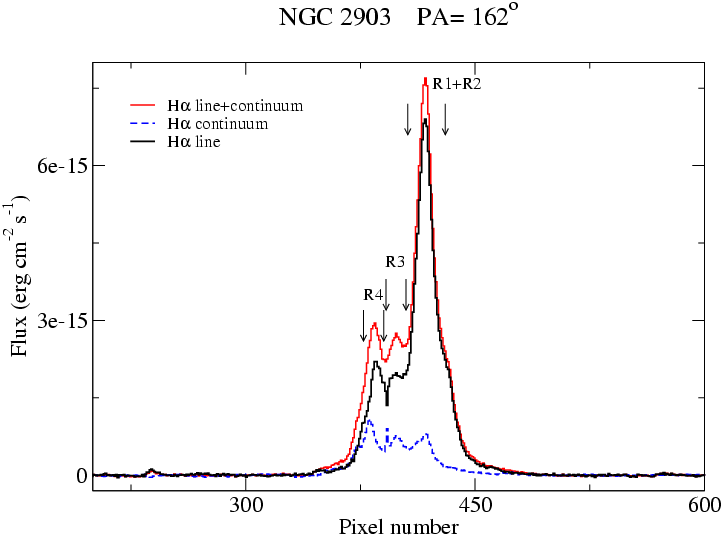}\\
\vspace{0.5cm}
\includegraphics[width=.41\textwidth,angle=0]{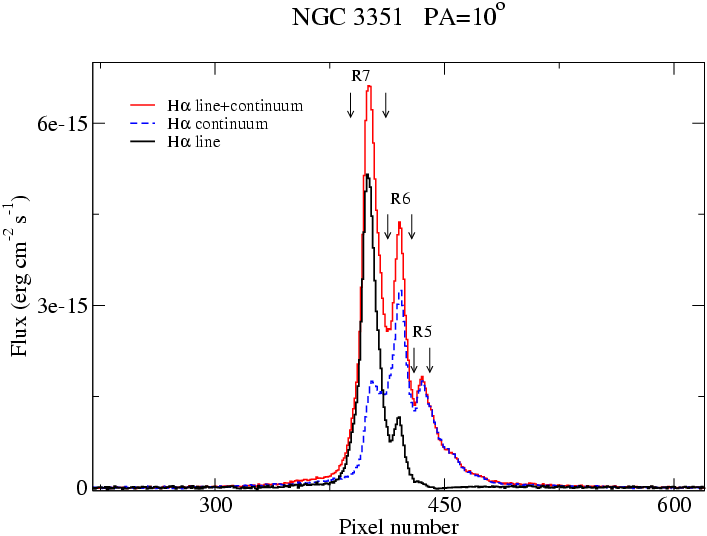}
\hspace{0.2cm}
\includegraphics[width=.41\textwidth,angle=0]{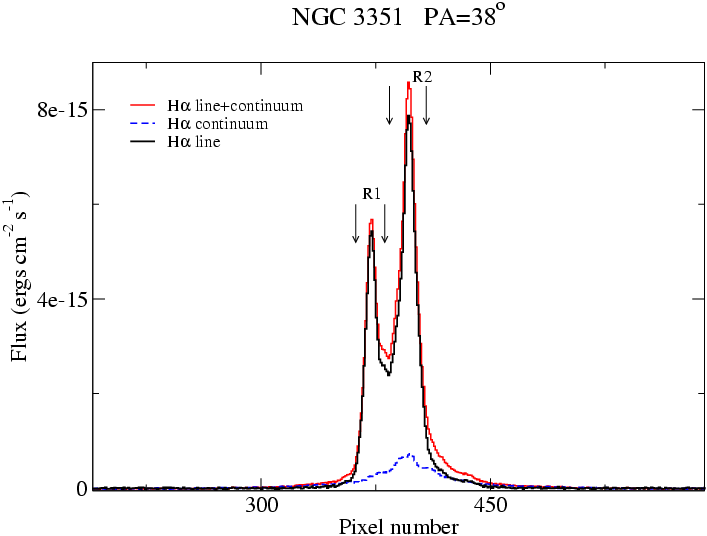}\\
\vspace{0.5cm}
\includegraphics[width=.41\textwidth,angle=0]{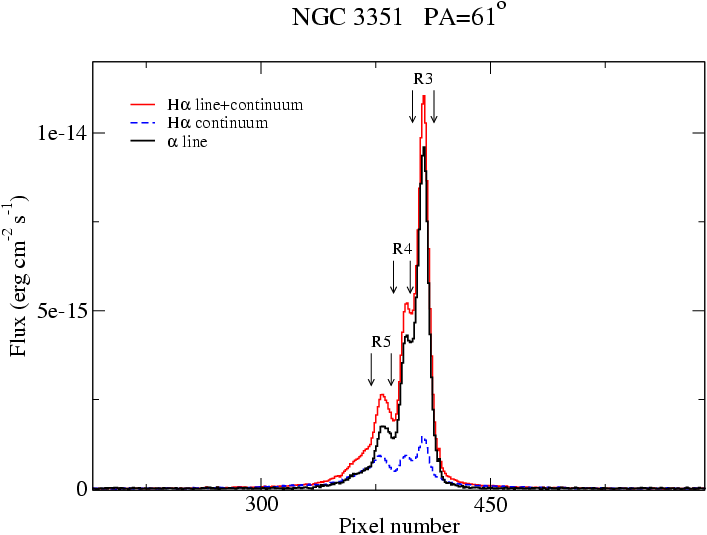}
\hspace{0.2cm}
\includegraphics[width=.41\textwidth,angle=0]{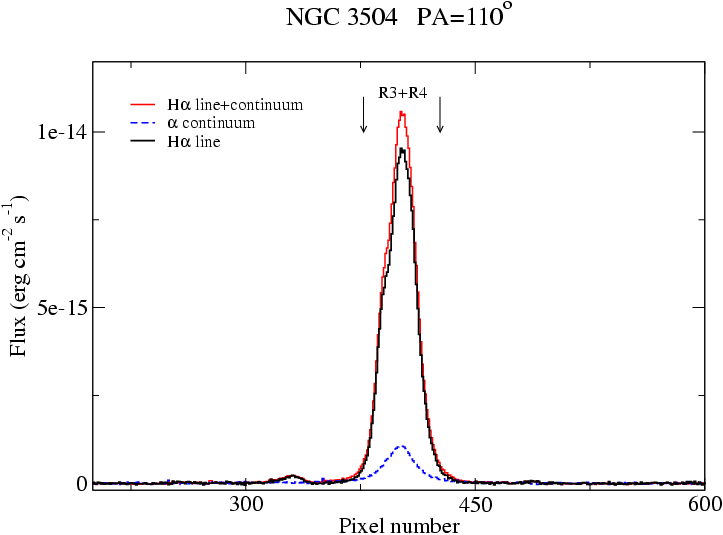}\\
\vspace{0.2cm}
\caption[]{H$\alpha$ profiles for the observed slit positions. Each figure
includes the name of the galaxy, the P.A. of the slit, the name of the observed
regions and their X,Y positions (see also Table 2).}
\label{profiles}
\end{figure*}

%%%%%%%%%%%%%%%%%%%%%%%%%%%%%%%%%%%%%%%%%%%%%%%%%%%%%%%%%%%%%%%%%%%%%%%%%%%%%%%

\section{Results}

\begin{figure*}
\centering
\vspace{0.5cm}
\includegraphics[width=.41\textwidth,angle=0]{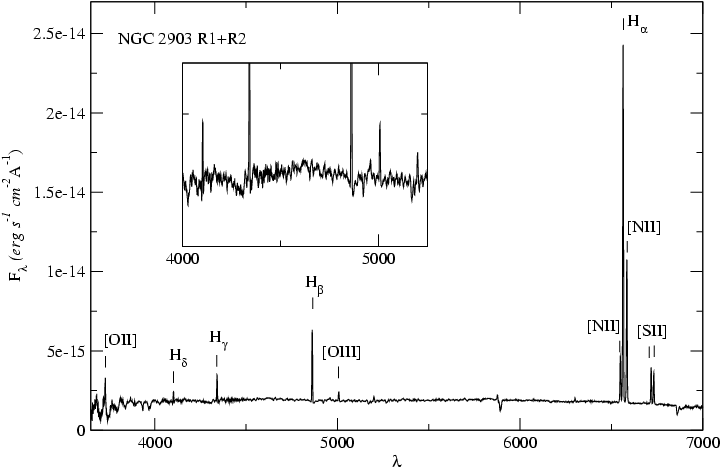}
\hspace{0.2cm}
\includegraphics[width=.41\textwidth,angle=0]{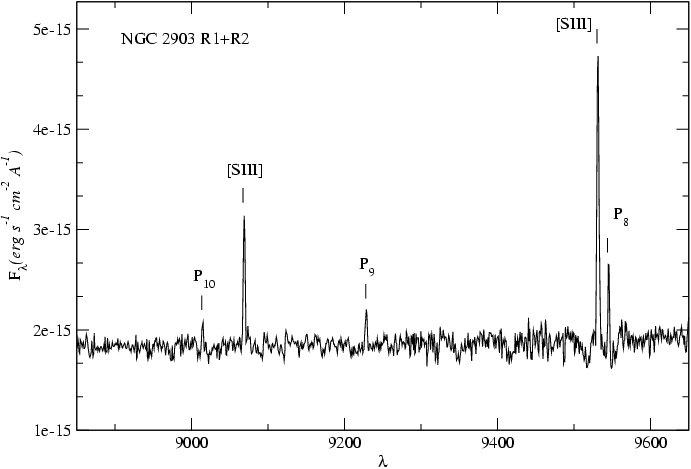}\\
%\hspace{0.2cm}
\vspace{0.9cm}
\includegraphics[width=.41\textwidth,angle=0]{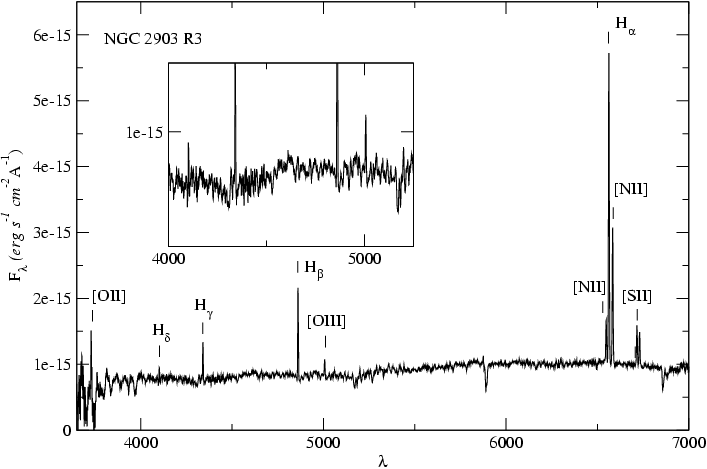}
\hspace{0.2cm}
\includegraphics[width=.41\textwidth,angle=0]{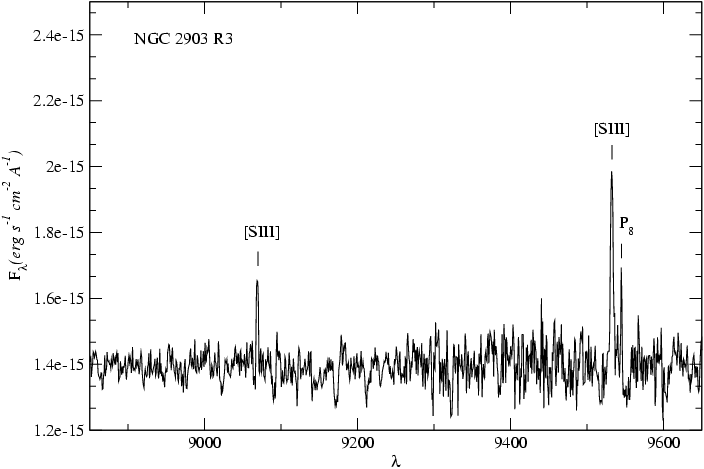}\\
\vspace{0.9cm}
\includegraphics[width=.41\textwidth,angle=0]{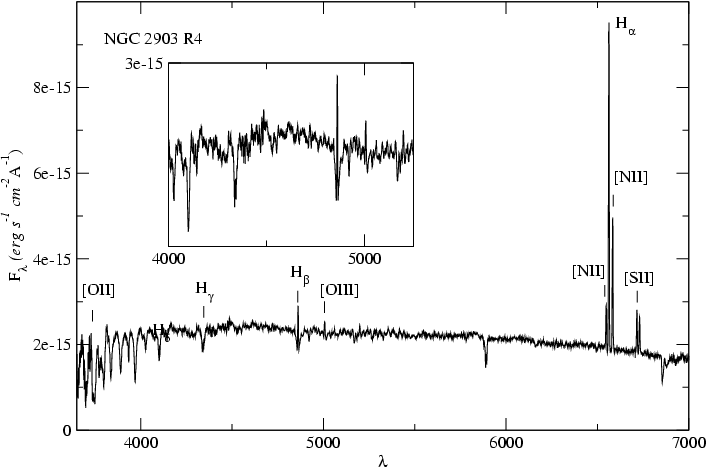}
\hspace{0.2cm}
\includegraphics[width=.41\textwidth,angle=0]{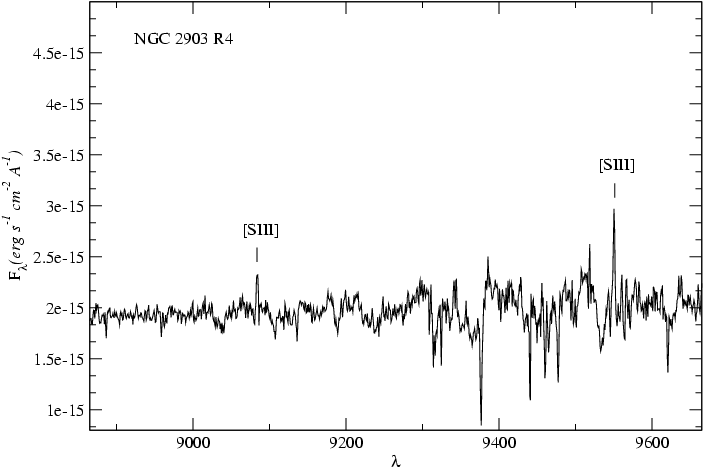}\\
\vspace{0.9cm}
\includegraphics[width=.41\textwidth,angle=0]{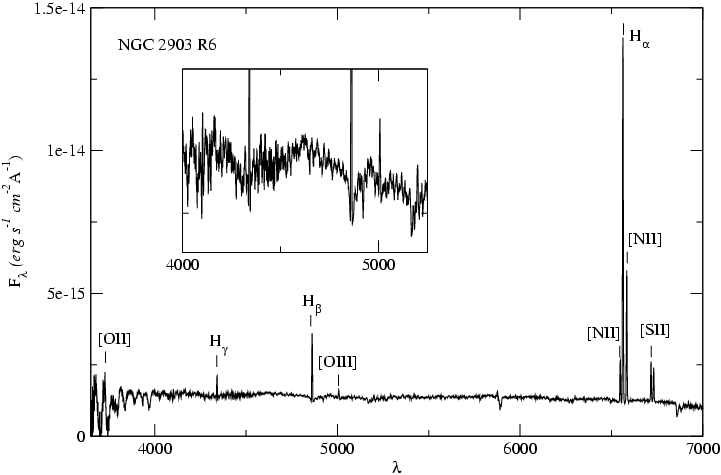}
\hspace{0.2cm}
\includegraphics[width=.41\textwidth,angle=0]{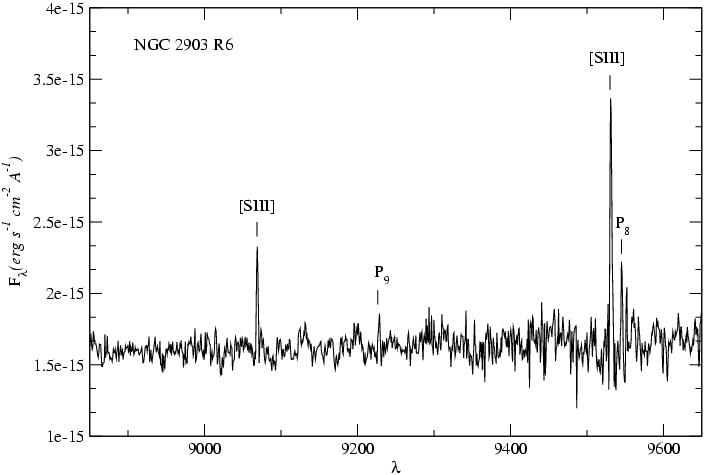}\\
\hspace{0.9cm}
\caption[]{Extracted blue (left) and red (right) spectra for the observed regions of NGC~2903. From top to bottom: R1+R2, R3, R4 and R6.}
\label{spectra1}
\end{figure*}

\begin{figure*}
\centering
\vspace{0.5cm}
\includegraphics[width=.41\textwidth,angle=0]{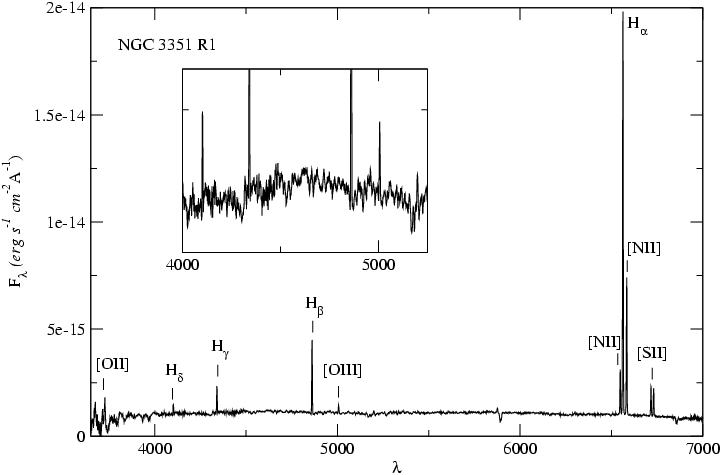}
\hspace{0.2cm}
\includegraphics[width=.41\textwidth,angle=0]{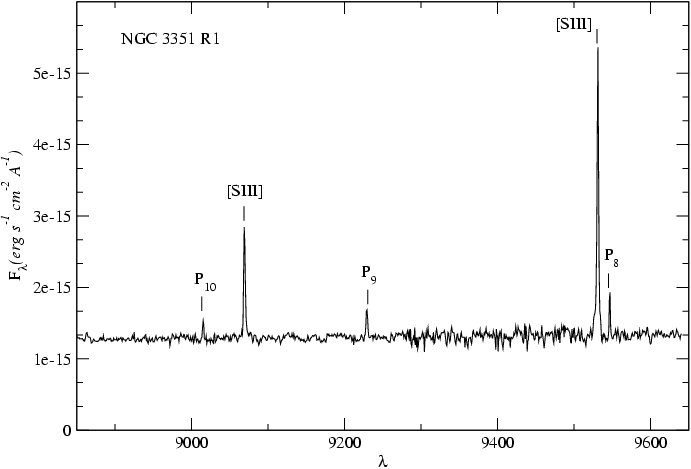}\\
%\hspace{0.2cm}
\vspace{0.9cm}
\includegraphics[width=.41\textwidth,angle=0]{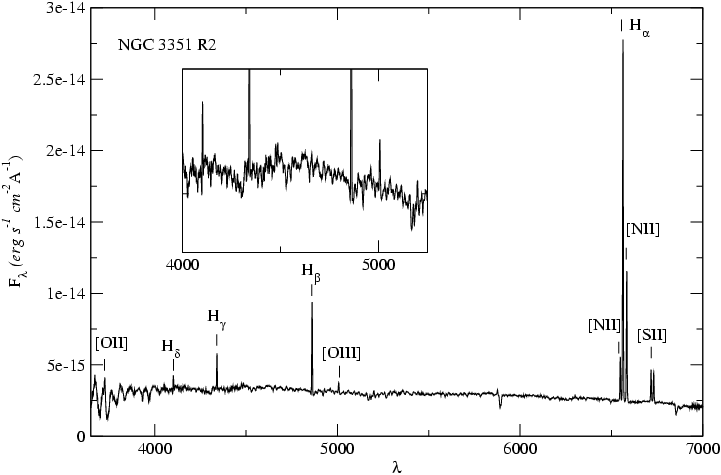}
\hspace{0.2cm}
\includegraphics[width=.41\textwidth,angle=0]{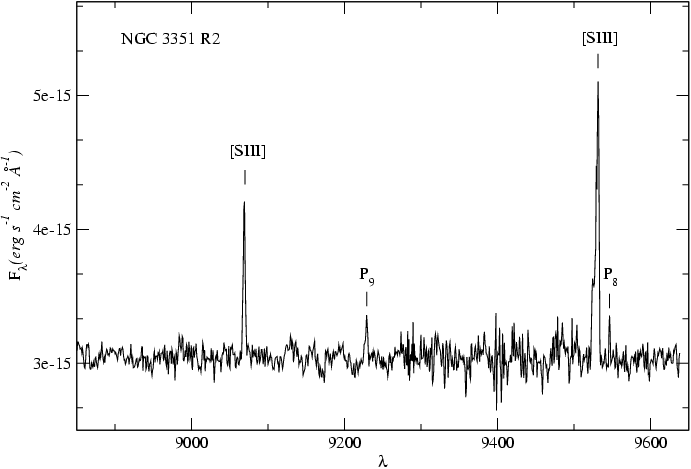}\\
\vspace{0.9cm}
\includegraphics[width=.41\textwidth,angle=0]{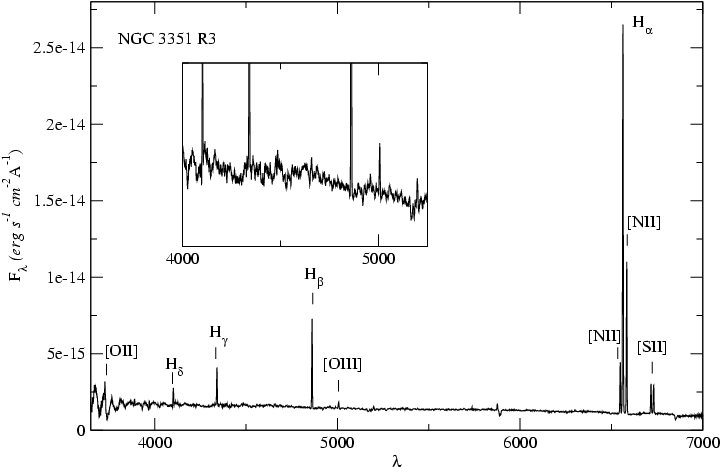}
\hspace{0.2cm}
\includegraphics[width=.41\textwidth,angle=0]{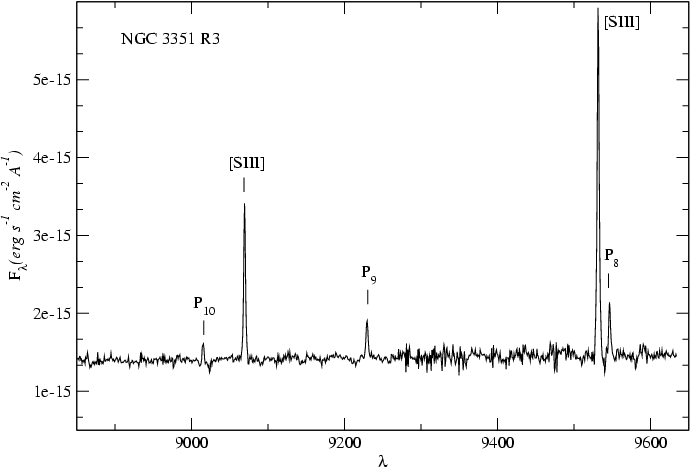}\\
\vspace{0.9cm}
\includegraphics[width=.41\textwidth,angle=0]{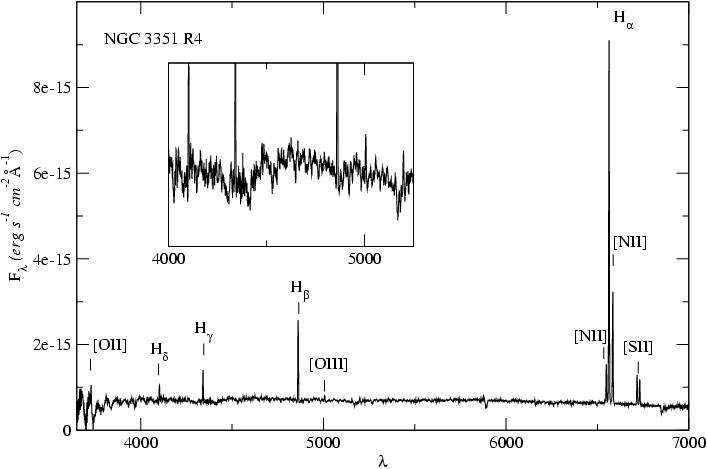}
\hspace{0.2cm}
\includegraphics[width=.41\textwidth,angle=0]{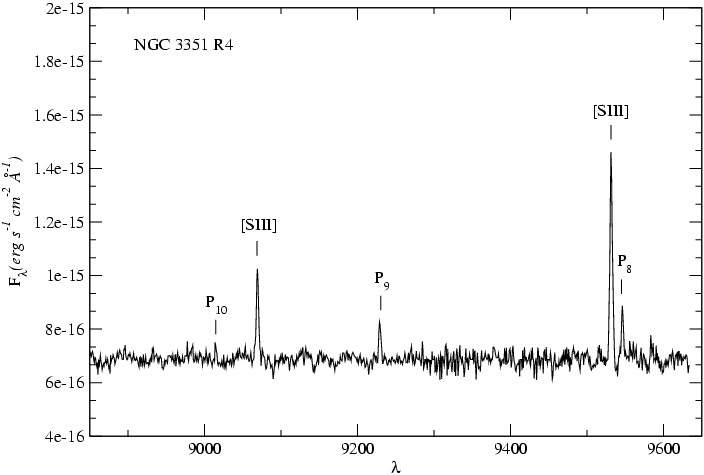}\\
\hspace{0.9cm}
\caption[]{Extracted blue (left) and red (right) spectra for the observed regions of NGC~3351. From top to bottom: R1, R2, R3 and R4.}
\label{spectra2}
\end{figure*}

Figure  \ref{profiles} shows the spatial distribution of the H$\alpha$ flux along the slit for the six different positions observed in the sample, a single one in the case of NGC~3504, two in NGC~2903 and three in NGC~3351. The regions that were extracted into 1-D spectra are delimited by arrows. The spectra corresponding to each of the identified regions are shown in Figures \ref{spectra1}, \ref{spectra2} and \ref{spectra3} for NGC~2903, NGC~3351 and NGC~3504 respectively.

%%%%%%%%%%%%%%%%%%%%%%%%%%%%%%%%%%%%%%%%%%%%%%%%%%%%%%%%%%%%%%%%%%%%%%%%%%%%%%%%
%                                                                              
%           Figuras 2-4: Espectros observados de cada region                                                                                                                 
%                                                                              
%%%%%%%%%%%%%%%%%%%%%%%%%%%%%%%%%%%%%%%%%%%%%%%%%%%%%%%%%%%%%%%%%%%%%%%%%%%%%%%%
%%%%%%%%%%%%%%%%%%%%%%%%%%%%%%%%%%%%%%%%%%%%%%%%%%%%%%%%%%%%%%%%%%%%%%%%%%%%%%%%

%%%%%%%%%%%%%%%%%%%%%%%%%%%%%%%%%%%%%%%%%%%%%%%%%%%%%%%%%%%%%%%%%%%%%%%%%%%%%%%%
%%%%%%%%%%%%%%%%%%%%%%%%%%%%%%%%%%%%%%%%%%%%%%%%%%%%%%%%%%%%%%%%%%%%%%%%%%%%%%%%

%%%%%%%%%%%%%%%%%%%%%%%%%%%%%%%%%%%%%%%%%%%%%%%%%%%%%%%%%%%%%%%%%%%%%%%%%%%%%%%%
\setcounter{figure}{3}
\begin{figure*}
\centering
\vspace{0.5cm}
\includegraphics[width=.41\textwidth,angle=0]{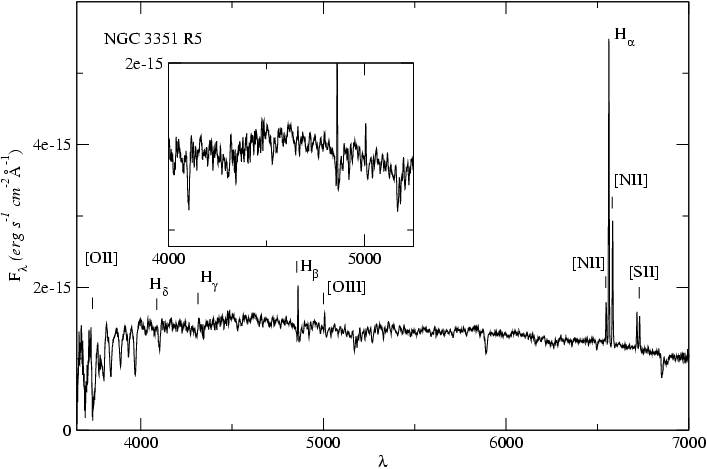}
\hspace{0.2cm}
\includegraphics[width=.41\textwidth,angle=0]{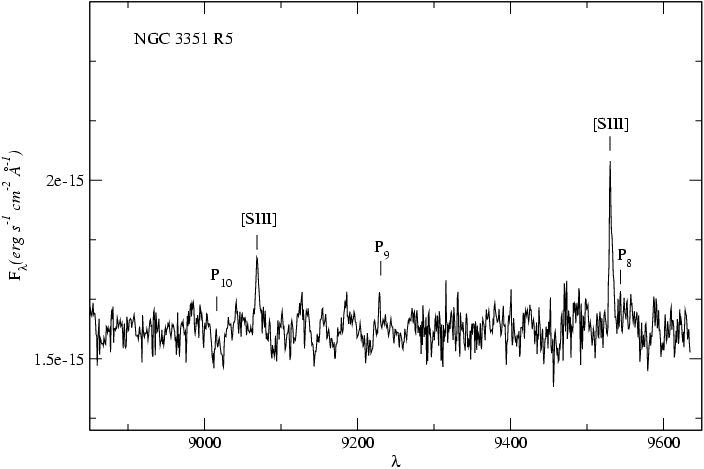}\\
\vspace{0.9cm}
\includegraphics[width=.41\textwidth,angle=0]{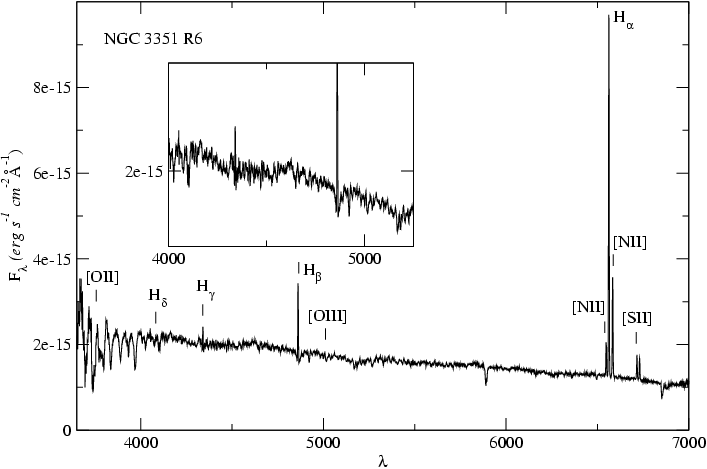}
\hspace{0.2cm}
\includegraphics[width=.41\textwidth,angle=0]{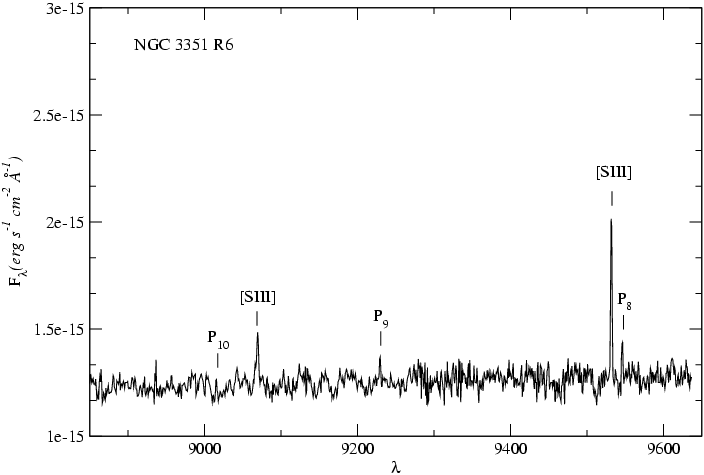}\\
\vspace{0.9cm}
\includegraphics[width=.41\textwidth,angle=0]{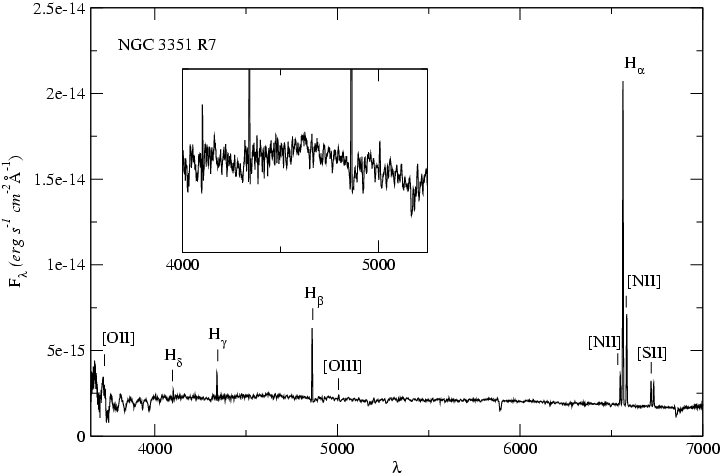}
\hspace{0.2cm}
\includegraphics[width=.41\textwidth,angle=0]{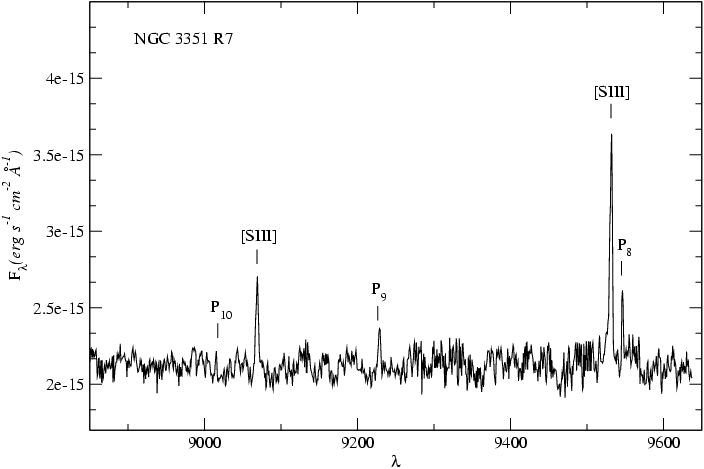}\\
\vspace{0.9cm}
\caption[]{({\it cont})Extracted blue (left) and red (right) spectra for the observed regions of NGC~3351. From top to bottom: R5, R6 and R7.}
%\label{spectra2}
\end{figure*}
%%%%%%%%%%%%%%%%%%%%%%%%%%%%%%%%%%%%%%%%%%%%%%%%%%%%%%%%%%%%%%%%%%%%%%%%%%%%%%%%

%%%%%%%%%%%%%%%%%%%%%%%%%%%%%%%%%%%%%%%%%%%%%%%%%%%%%%%%%%%%%%%%%%%%%%%%%%%%%%%%
\setcounter{figure}{4}
\begin{figure*}
\centering
\vspace{0.5cm}
\includegraphics[width=.41\textwidth,angle=0]{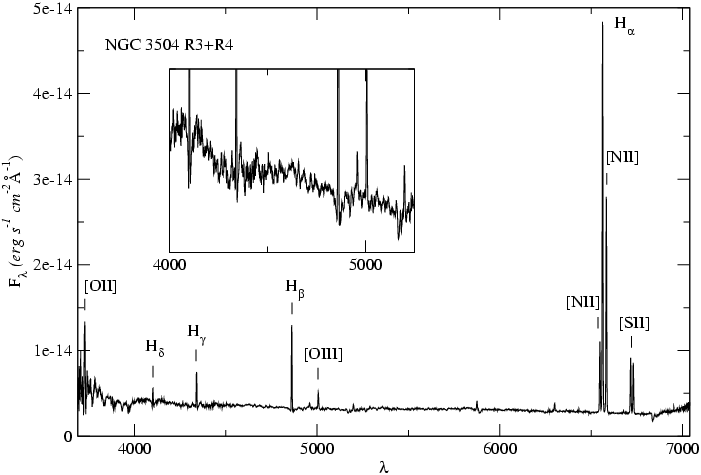}
\hspace{0.2cm}
\includegraphics[width=.41\textwidth,angle=0]{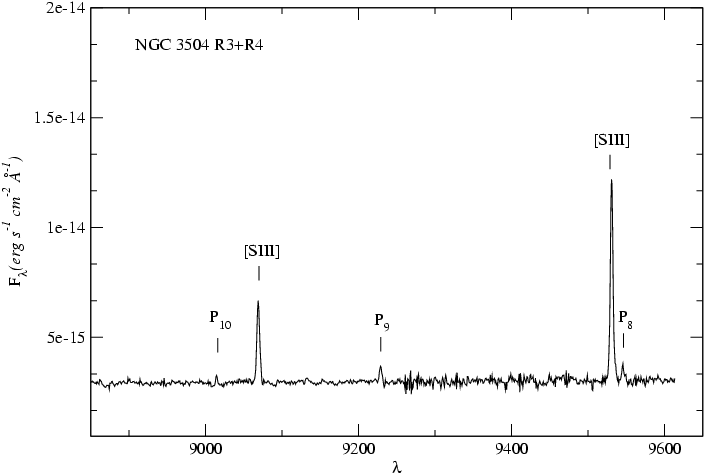}
%\hspace{0.2cm}
\vspace*{0.9cm}
\caption[]{Extracted blue (left) and red (right) spectra for regions R3 + R4 of NGC~3504.}
\label{spectra3}
\end{figure*}

\subsection{Underlying population}
\begin{figure*}
\centering
\includegraphics[width=.41\textwidth,height=0.20\textwidth,angle=0]{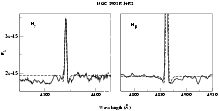}\\
\vspace{0.2cm}
\includegraphics[width=.41\textwidth,height=0.20\textwidth,angle=0]{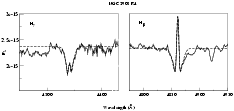}
\caption[]{Examples of the fitting procedure used to correct the Balmer emission line intensities for underlying absorption.}
\label{underlying}
\end{figure*}

The presence of underlying Balmer stellar absorptions is clearly evident in the blue spectra of the observed regions (see Figures \ref{spectra1}, \ref{spectra2} and \ref{spectra3}) and complicates the measurements. A two-component -- emission and absorption -- gaussian fit was performed in order to correct the Balmer emission lines for this effect. An example of this procedure can be seen in Figure \ref{underlying}. The equivalent widths (in \aa mstrongs) of the gaussian absorption components resulting from the fits are given in Table 4 together with the ratio between the line flux measured after subtraction of the absorption component and the line flux measured without any correction and using a pseudo-continuum placed at the bottom of the line. This factor provides a value for the final correction to the measured fluxes in terms of each line flux. In regions R4 of NGC~2903 and R5 of NGC~3351 the H$\delta$ line is seen only in absorption. In region R5 of NGC~3351 also H$\gamma$ is seen only in absorption. No fitting was performed for these lines, hence no correction is listed for them in Table 4.
In the case of the HeI and Paschen no prominent absorption line wings are observed that allow the fitting of an absorption component as it was done in the case of the Balmer lines. These lines were measured with respect to a local continuum placed at the their base, which partially corrects by underlying absorption.

%%%%%%%%%%%%%%%%%%%%%%%%%%%%%%%%%%%%%%%%%%%%%%%%%%%%%%%%%%%%
%                                                                             
%             FIGURA  : Ajuste de las absorciones                        
%                                                                             
%%%%%%%%%%%%%%%%%%%%%%%%%%%%%%%%%%%%%%%%%%%%%%%%%%%%%%%%%%%%%%%%%%%%%%%%%%%%%%%
%\vspace*{2.5cm}

%%%%%%%%%%%%%%%%%%%%%%%%%%%%%%%%%%%%%%%%%%%%%%%%%%%%%%%%%%%%
%                                                                             
%             TABLA 4  : Equivalent widths of Balmer absorption lines and percentage correction to the line                       
%                                                                             
%%%%%%%%%%%%%%%%%%%%%%%%%%%%%%%%%%%%%%%%%%%%%%%%%%%%%%%%%%%%%%%%%%%%%%%%%%%%%%%

\begin{table*}
\centering
 \begin{minipage}{148mm}
\label{balmer}
 \caption{Equivalent widths of Balmer absorption lines for the observed CNSFR}
 \begin{tabular}{@{}lcccccccc@{}}
\hline
\hline
Galaxy & Region & H$\delta$ (\AA ) & Corr. factor & H$\gamma$ (\AA ) & Corr. factor & H$\beta$ (\AA )& Corr. factor\\
\hline
NGC~2903 & R1+R2 & 2.3 & 1.15 & 3.1 & 1.17 & 2.9 & 1.09 \\
                 & R3       & 2.8 & 1.17 & 3.6 & 1.21 & 3.5 & 1.09 \\
                 & R4       & 4.5 &   --   & 3.8 & 1.08 & 4.8 & 1.24 \\
                 & R6       & 3.0 & 1.04 & 2.6 & 1.06 & 3.9 & 1.18 \\
NGC~3351 & R1       & 2.1 & 1.16 & 1.6 & 1.14 & 3.0 & 1.08 \\
                 & R2       & 2.4 & 1.30 & 1.2 & 1.09 & 3.2 & 1.06 \\
                 & R3       & 2.5 & 1.16 & 1.2 & 1.07 & 1.2 & 1.03 \\
                 & R4       & 2.1 & 1.04 & 1.2 & 1.08 & 2.5 & 1.07 \\
                 & R5       & 3.7 & --     & 2.5 &  --    & 3.7 & 1.27 \\
                 & R6       & 2.4 & 1.32 & 2.0 & 1.04 & 3.5 & 1.13 \\
                 & R7       & 2.4 & 1.26 & 2.1 & 1.17 & 3.4 & 1.09 \\
NGC~3504 & R3+R4 & 4.5 & 1.25 & 3.6 & 1.18 & 4.4 & 1.10 \\
\hline
\end{tabular}
\end{minipage}
\end{table*}

\subsection{Line intensity measurements}

Emission line fluxes were measured on the extracted spectra using the IRAF SPLOT software package, by integrating the line intensity over a local fitted continuum. The errors in the observed line fluxes have been calculated from the expression $\sigma_{l}$ = $\sigma_{c}$N$^{1/2}$[1 + EW/(N$\Delta$)]$^{1/2}$, where $\sigma_{l}$ is the error in the line flux, $\sigma_{c}$ represents the standard deviation in a box near the measured emission line and stands for the error in the continuum placement, N is the number of pixels used in the measurement of the line flux, EW is the line equivalent width, and $\Delta$ is the wavelength dispersion in \aa ngstroms per pixel. The first term represents the error in the line flux introduced by the uncertainty in the placement of the continuum, while the second one scales the signal-to-noise in the continuum to the line \citep{1994ApJ...437..239G}.

\begin{figure*}
\centering
\includegraphics[width=.30\textwidth,angle=0]{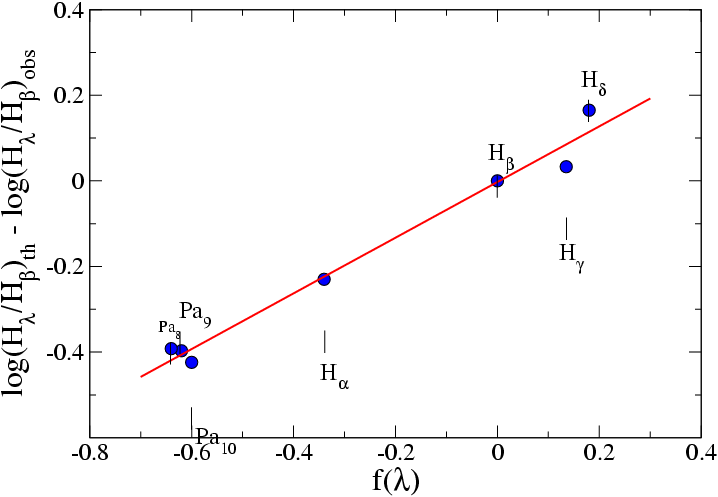}
\caption[]{Reddening determination for region R1+R2 in NGC~2903. The tight relation found shows the goodness of the correction to the Balmer emission lines by the underlying absorption continuum.}
\label{reddening}
\end{figure*}

The Balmer emission lines were corrected for the underlying absorption as explained above. Then the logarithmic extinction at H$\beta$, c(H$\beta$) was calculated from the Balmer line decrements assuming the Balmer line theoretical values for  case B recombination \cite{1971MNRAS.153..471B} for a temperature of 6000 K, as expected for high metallicity regions, and an average extinction law \cite{1972ApJ...172..593M}.
 An example of this procedure is shown in Figure \ref{reddening} for region R1+R2 of NGC~2903, where the logarithm of the quotient between observed and theoretical Balmer decrements is represented against the logarithmic extinction at the Balmer line wavelengths, f($\lambda$).  The tight relation found for a baseline from the Paschen lines to H$\delta$ can be taken as evidence of the reliability of our subtraction procedure for this region. This is also the case for most regions, although in some cases the H$\gamma$/H$\beta$ ratio lies above or below the reddening line, implying that the subtraction procedure is not that good for this line. This is not surprising since the presence of the G-band and other metal features in the left wing of the line, makes the fitting less reliable than for the other Balmer lines. We have used all the available Balmer and Paschen-to-Balmer ratios, although due to the uncertainties in the emission line intensities of the higher order Balmer lines, which are difficult to estimate (in fact, in some regions they are seen only in absorption), these ratios have been given a lower weight in the fit. This almost amounts to deriving the values of c(H$\beta$) from the H$\alpha$/H$\beta$ ratio and checking for consistency against the values measured for the Paschen lines, which are much less affected by underlying absorption due the smaller contribution to the continuum from main sequence AF stars. The errors in c(H$\beta$) have in fact been derived from the measured errors in the H$\alpha$ and H$\beta$ lines, since in any weighted average these two lines have by far the largest weight.  

\begin{figure*}
\centering
\includegraphics[width=.30\textwidth,angle=0]{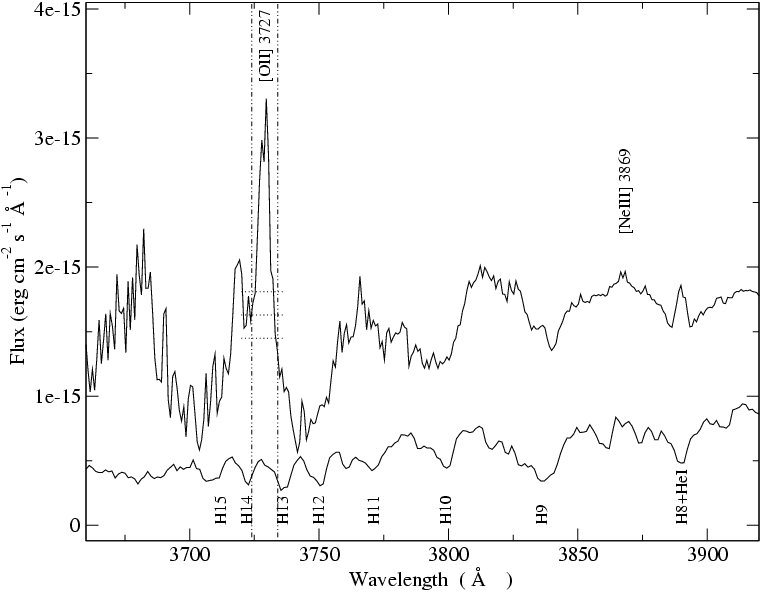}
\caption[]{Blue spectrum of region R1+R2 in NGC~2903 showing the location of the [OII] $\lambda\lambda$ 3727,29 \AA\ lines. The spectrum below corresponds to a metal rich globular cluster in M~31 which can be compared with the underlying stellar population in the region. The way in which the [OII] line has been measured by placing a local pseudo-continuum at its base is shown. The different continuum placementes used for the computation of the errors are shown by horizontal lines. }
\label{medidas1}
\end{figure*}

The blue region of the spectrum near the Balmer discontinuity is dominated by absorption lines which cause a depression of the continuum and difficult the measurement of the [OII] lines at $\lambda$ 3727 \AA\. None of them however is at the actual wavelength of the [OII] line. This can be seen in Figure \ref{medidas1} where we show the spectrum of region R1+R2 in NGC~2903 compared to that corresponding to a globular cluster of M~31, 337-068, of relatively high metallicity (Barmby et al.~2000; Mike Beasley, private communication).  We have measured the line using a local continuum at its base as shown in the figure.  The different continuum placements used for computing  the error are shown by horizontal dotted lines in the figure.

%%%%%%%%%%%%%%%%%%%%%%%%%%%%%%%%%%%%%%%%%%%%%%%%%%%%%%%%%%%%
%                                                                             
%             FIGURA  : Reddening determination for R1+R2 in NGC 2903                        
%                                                                             
%%%%%%%%%%%%%%%%%%%%%%%%%%%%%%%%%%%%%%%%%%%%%%%%%%%%%%%%%%%%%%%%%%%%%%%%%%%%%%%

Once the reddening constant was found, the measured line intensities relative to the H$\beta$ line were corrected for interstellar reddening according to the assumed reddening law. The errors in the reddening corrected line intensities have been derived by means of error propagation theory. 
Measured and reddening corrected emission line fluxes, together with their corresponding errors,  are given in Tables \ref{intensities1}, \ref{intensities2} and \ref{intensities3} for the observed CNSFR in NGC~2903, NGC~3351 and NGC~3504 respectively.  Balmer emission lines are corrected for underlying absorption. Also given in the tables are the assumed reddening law, the  H$\beta$ intensity underlying  absorption and extinction corrected,  the H$\beta$ equivalent 
width, also corrected for absorption, and the reddening constant. 

%%%%%%%%%%%%%%%%%%%%%%%%%%%%%%%%%%%%%%%%%%%%%%%%%%%%%%%%%%%%
%                                                                             
%             FIGURA  : Measurement of the [OII] lines            
%                                                                             
%%%%%%%%%%%%%%%%%%%%%%%%%%%%%%%%%%%%%%%%%%%%%%%%%%%%%%%%%%%%%%%%%%%%%%%%%%%%%%%

\section{Chemical abundances} 

Electron densities for each observed region have been derived from the  [S{\sevensize II}] $\lambda\lambda$ 6717, 6731 \AA\ line ratio, following standard methods \citep[e.g.][]{1989agna.book.....O}. They were found to be, in all cases, 
$\le$ 600 cm$^{-3}$, higher than those usually derived in disk HII regions, but still below the critical value for collisional de-excitation.

The low excitation of the regions, as evidenced by the weakness of the [OIII] $\lambda$ 5007 \AA\ line (see left panels of Figures \ref{spectra1}, \ref{spectra2} and \ref{spectra3}), precludes the detection and measurement of the auroral [OIII] $\lambda$ 4363 \AA\ necessary for the derivation of the electron temperature. It is therefore impossible to obtain a direct determination of the oxygen abundances. Empirical calibrations have to be used instead. 

Different calibrators for strong emission lines have been proposed in the literature for different kinds of objects involving different chemical elements, among others, the oxygen abundance parameter R$_{23}$ $ \equiv $ O$_{23}$ \citep{1979MNRAS.189...95P}, the nitrogen N2 parameter \citep{2002MNRAS.330...69D} and the sulphur abundance parameter  S$_{23}$ \citep{2000MNRAS.312..130D}. Also, calibrators involving a combination of emision lines of two elements have been proposed, such as [OIII]/[NII] \citep{1979A&A....78..200A}, [ArIII]$\lambda$ 7135 \AA /[OIII] $\lambda$ 5007 \AA\ and [SIII]$\lambda$ 9069 \AA /[OIII] $\lambda$ 5007 \AA\ \citep{2006A&A...454L.127S}. Our CNSFR, show very weak lines of [OIII] which are measured with large errors. This is taken as evidence for high metallicity in these regions. This, in turn, may imply values of N/O larger than solar, due to the chemical evolution of the regions themselves and the increasing production of secondary nitrogen. Hence, we have considered the use of the N2 parameter unreliable for this kind of objects. On the other hand, the [SIII] lines are seen to be strong as compared to the [OIII] lines (see right panels of Figures \ref{spectra1}, \ref{spectra2} and \ref{spectra3}). Recently, the combination of both the oxygen abundance, O$_{23}$, and sulphur abundance, S$_{23}$, parameters has been claimed to be a good metallicity indicator for high metallicity HII regions \citep{2000MNRAS.312..130D,2005MNRAS.361.1063P}. From now onwards we will call this parameter SO$_{23}$, and is defined as:
\[
SO_{23} = \frac{S_{23}}{O_{23}} = \frac{I([SII]\lambda 6716,31)+I([SIII]\lambda 9069,9532)}
{I([OII]\lambda 3727,29)+I([OIII]\lambda 4959,5007)} \]

\begin{figure*}
\vspace*{1cm}
\centering
\includegraphics[width=.41\textwidth,angle=0]{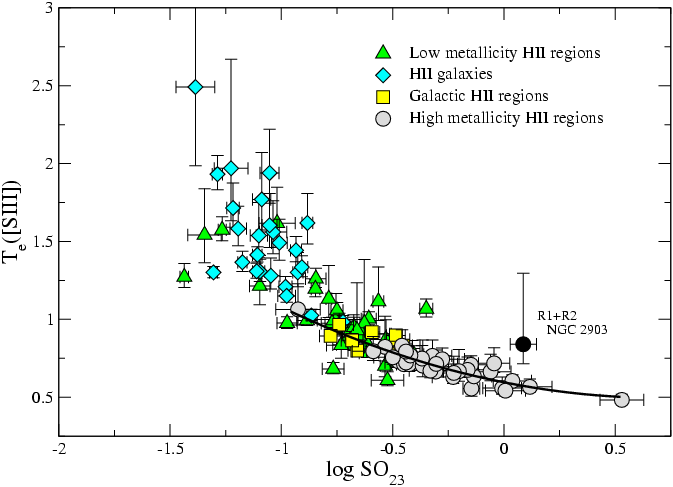}
\caption[]{Empirical calibration of the [SIII] electron temperature as a function of the abundance parameter SO$_{23}$ defined in the text. The solid line represents a quadratic fit to the high metallicity HII region data. References for the data are given in the text.}
\label{te[SIII]cal}
\end{figure*}

This parameter is similar to the S$_3$O$_3$ proposed by \cite{2006A&A...454L.127S} but is, at first order, independent of geometrical (ionization parameter) effects. The amount of available data on sulphur emission lines is increasingly growing, especially in the high metallicity regime. This makes possible for the first time to calibrate the electron temperature of [SIII] in terms of the SO$_{23}$ parameter. To perform this calibration we have compiled all the data so far at hand with sulphur emission line data both for the auroral and nebular lines at $\lambda$ 6312 \AA\ and $\lambda\lambda$ 9069,9532 \AA\ respectively. The sample comprises data on galactic (Garc\'\i a-Rojas 2006) and extragalactic \citep{1987MNRAS.226...19D,1988MNRAS.235..633V,1993MNRAS.260..177P,1997ApJ...489...63G,
1994ApJ...437..239G,1995ApJ...439..604G,2000MNRAS.318..462D,2002MNRAS.329..315C,2003ApJ...591..801K, 2004ApJ...615..228B,2005A&A...441..981B} HII regions and HII galaxies \citep{1993ApJ...411..655S,1994ApJ...431..172S,2003MNRAS.346..105P,2006MNRAS.372..293H,
Hageleetal.2007}.
 The data on extragalactic HII regions has been further split into low and high metalicity HII regions according to the criterion of \cite{2000MNRAS.312..130D}, {\it i.e.} log$O_{23} \leq$ 0.47 and -0.5$\leq log\,S_{23} \leq$ 0.28 
implies oversolar abundances. For all the regions the [SIII] electron temperature has been derived from the ratio between the auroral and the nebular sulphur lines, using a five-level atom program \citep{1995PASP..107..896S} and the collisional strengths from \cite{1999ApJ...526..544T}, through the task TEMDEN as implemented in the IRAF package. The calibration is shown in Figure \ref{te[SIII]cal} together with the quadratic fit to the high metallicity HII region data:
\[
t_e([SIII]) = 0.596 - 0.283 log SO_{23} + 0.199 (log SO_{23})^2
\]

\begin{figure*}
\centering
\vspace*{0.5cm}
\includegraphics[width=.41\textwidth,angle=0]{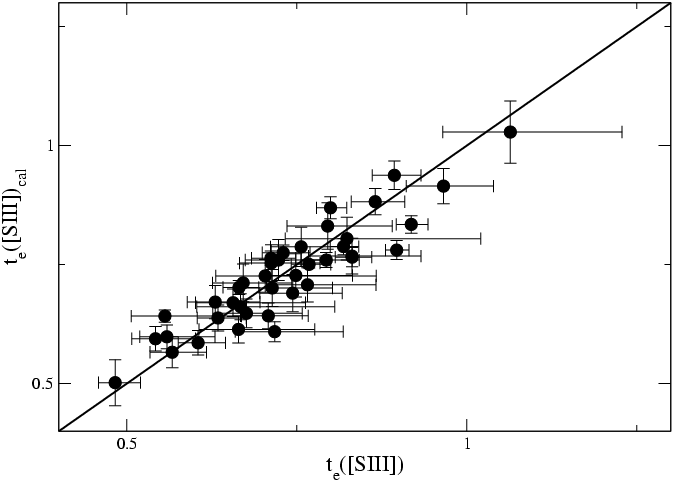}
\hspace{0.2cm}
\includegraphics[width=.41\textwidth,angle=0]{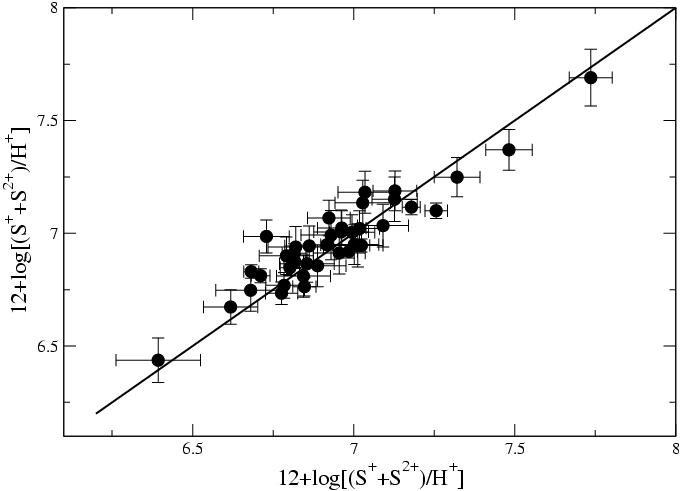}

\caption[]{Comparison between the electron temperatures (left) and ionic abundances for sulphur (right) for the high metallicity HII region sample, as derived from measurements of the auroral sulphur line $\lambda$ 6312 \AA\ (abscissa) and from our derived calibration (ordinate). The solid line represents the one-to-one relation.}
\label{comparison}
\end{figure*}

Figure \ref{comparison} shows the comparison between the electron temperatures and ionic sulphur abundances derived from measurements of the auroral sulphur line $\lambda$ 6312 \AA\ and our derived calibration.  The S$^+$/H$^+$ and S$^{++}$/H$^+$ ionic ratios have been derived from: 

$
\begin{array}{ll}
$$12+log\frac{S^+}{H^+}=$$ & $$log\left( \frac{I(6717)+I(6731)}{I(H\beta} \right) +$$   \\ 
                                       &  $$ 5.423 +\frac{0.929}{t}-0.28 \cdot logt $$\\ 
$$12+log\frac{S^{2+}}{H^+}=$$ & $$ log\left( \frac{I(9069)+I(9532)}{I(H\beta} \right) + $$ \\ 
                                       & $$5.8 +\frac{0.771}{t}-0.22 \cdot logt$$
\end{array} 
$

These expressions have been derived by performing appropriate fittings to the IONIC task results following the functional form given in \cite{1992MNRAS.255..325P}.
The assumption has been made that T$_e$[SIII] $\simeq$T$_e$([SII]) in the observed regions. This assumption seems to be justified in view of the results presented by \cite{2005A&A...441..981B}.

We have used the calibration above to derive t$_{e}$([SIII]) for our observed CNSFR. In all cases the values of log$O_{23}$ are inside the range used in performing the calibration, thus requiring no extrapolation of the fit.
%Based on the residuals of the calibration, an average uncertainty of $\pm$300  K has been adscribed to the resulting temperatures. 
These temperatures, in turn, have been used to derive the S$^+$/H$^+$ and S$^{++}$/H$^+$ ionic ratios.
The derived T$_e$ and ionic abundances for sulphur are given in Table \ref{ionic-sulphur}, together with measured values of the electron density. Temperature and abundance errors are formal errors calculated from the measured line intensity errors applying error propagation formulae,  without assigning any error to the temperature calibration itself. 
%The whole procedure carries much more uncertainty that this, and the values derived should only be considered as estimates.

%%%%%%%%%%%%%%%%%%%%%%%%%%%%%%%%%%%%%%%%%%%%%%%%%%%%%%%%%%%%%%
%
%   TABLA 5 : REDDENING CORRECTED LINE INTENSITIES IN THE OBSERVED CNSFR OF NGC 2903
%
%%%%%%%%%%%%%%%%%%%%%%%%%%%%%%%%%%%%%%%%%%%%%%%%%%%%%%%%%%%%%%

\newpage
\onecolumn
\landscape

\begin{table}
%\centering
\vspace{3cm}

\caption[]{Reddening corrected emission line intensities for the CNSFR in NGC~2903}
\begin{tabular}{l c c c c c c c c c c}
\hline
Region & & &\multicolumn{2}{c}{R1+R2} & \multicolumn{2}{c}{R3} & \multicolumn{2}{c}{R4} & \multicolumn{2}{c}{R6}\\
            & &  &                &                  &            &                  &                  &                &              &    \\
 $\lambda$ (\AA ) & Line   & f$_\lambda$& F$_\lambda$& I$_\lambda$ &F$_\lambda$& I$_\lambda$&F$_\lambda$& I$_\lambda$&F$_\lambda$& I$_\lambda$ \\
            & &  &                &                  &            &                  &                  &                &              &     \\
 
3727 & [OII]      & 0.271 & 267$\pm$27  & 428$\pm$44 &  333 $\pm$ 33     &  430 $\pm$ 45  &  421 $\pm$ 42 & 841 $\pm$ 88 &   198 $\pm$ 20 &  310  $\pm$ 31 \\ 
4102 & H$_\delta$ & 0.188 &  179$\pm$11 &  248$\pm$16 &  235$\pm$16 &  280$\pm$19 &     --      &     --      &  145$\pm$9  &  197$\pm$13  \\
4340 & H$_\gamma$ & 0.142 &  432$\pm$13 &  552$\pm$17 &  553$\pm$18 &  632$\pm$20 &  164$\pm$17 &  235$\pm$24 &  298$\pm$10 &  377$\pm$13  \\
4686 & HeII       & 0.045 &    5$\pm$2  &    6$\pm$3  &     --      &     --      &     --      &     --      &  11$\pm$4 &  12$\pm$4  \\
4861 & H$_\beta$  & 0.000 & 1000$\pm$16 & 1000$\pm$16 & 1000$\pm$21 & 1000$\pm$21 & 1000$\pm$27 & 1000$\pm$28 & 1000$\pm$13 & 1000$\pm$13  \\ 
4959 & [OIII]     &-0.024 &   43$\pm$6  &   41$\pm$5  &   78$\pm$5  &   76$\pm$5  &  132$\pm$84 &  125$\pm$79 &   41$\pm$3  &   40$\pm$3   \\ 
5007 & [OIII]     &-0.035 &  131$\pm$5  &  123$\pm$5  &  229$\pm$13 &  222$\pm$12 &  391$\pm$25 &  357$\pm$23 &  122$\pm$9  &  115$\pm$8   \\
5199 & [NI]       &-0.078 &   56$\pm$8  &   49$\pm$7  &   65$\pm$26 &   60$\pm$24 &  139$\pm$45 &  114$\pm$37 &   60$\pm$15 &   53$\pm$13  \\
5876 & HeI        &-0.209 &   79$\pm$14 &   55$\pm$10 &   97$\pm$26 &   80$\pm$21 &     --      &    --       &   97$\pm$17 &   69$\pm$12  \\
6300 & [OI]       &-0.276 &   39$\pm$6  &   24$\pm$3  &   54$\pm$13 &   41$\pm$10 &   56$\pm$14 &   28$\pm$7  &   40$\pm$13 &   25$\pm$8   \\
6312 & [SIII]     &-0.278 &    7$\pm$1  &    4$\pm$1  &     --      &     --      &     --      &    --       &     --      &    --        \\
6548 & [NII]      &-0.311 &  675$\pm$11 &  394$\pm$12 &  536$\pm$16 &  400$\pm$17 &  866$\pm$62 &  391$\pm$31 &  529$\pm$17 &  318$\pm$12  \\
6563 & H$_\alpha$ &-0.313 & 4897$\pm$45 & 2850$\pm$26 & 3826$\pm$43 & 2850$\pm$32 & 6341$\pm$50 & 2850$\pm$22 & 4759$\pm$38 & 2850$\pm$23  \\
6584 & [NII]      &-0.316 & 2082$\pm$19 & 1206$\pm$33 & 1679$\pm$47 & 1247$\pm$54 & 2666$\pm$66 & 1189$\pm$52 & 1707$\pm$26 & 1017$\pm$27  \\
6678 & HeI        &-0.329 &    8$\pm$3  &    4$\pm$2  &     --      &     --      &     --      &     --      &     --      &     --       \\
6717 & [SII]      &-0.334 &  498$\pm$14 &  280$\pm$11 &  472$\pm$28 &  345$\pm$24 &  811$\pm$34 &  346$\pm$20 &  528$\pm$15 &  306$\pm$11  \\
6731 & [SII]      &-0.336 &  432$\pm$16 &  242$\pm$11 &  391$\pm$24 &  285$\pm$20 &  699$\pm$31 &  297$\pm$18 &  450$\pm$16 &  260$\pm$11  \\
8863 & P11        &-0.546 &   36$\pm$8  &   14$\pm$3  &     --      &     --      &     --      &     --      &     --      &     --       \\
9016 & P10        &-0.557 &   46$\pm$7  &   18$\pm$3  &     --      &     --      &     --      &     --      &   53$\pm$14 &   21$\pm$6   \\
9069 & [SIII]     &-0.561 &  157$\pm$10 &   60$\pm$5  &  103$\pm$9  &   61$\pm$6  &  220$\pm$33 &   53$\pm$9  &  137$\pm$11 &   55$\pm$5   \\
9230 & P9         &-0.572 &   61$\pm$8  &   23$\pm$3  &     --      &     --      &     --      &     --      &   87$\pm$16 &   34$\pm$6   \\
9532 & [SIII]     &-0.592 &  393$\pm$16 &  141$\pm$9  &  337$\pm$16 &  193$\pm$15 &  508$\pm$53 &   112$\pm$14 &  450$\pm$29 &  171$\pm$13  \\
9547 & P8         &-0.593 &  128$\pm$14 &   46$\pm$5  &  128$\pm$17 &   73$\pm$11 &  305$\pm$58 &   67$\pm$14 &  192$\pm$20 &   73$\pm$8   \\
          
\hline

c(H$\beta$)  & & & 0.75$\pm$0.04 & & 0.41$\pm$0.05 & & 1.11$\pm$0.05 & & 0.71$\pm$0.03 &  \\
I(H$\beta$)$^a$        & & & & 13.7$\pm$0.22 & & 1.80$\pm$0.04 & & 7.76$\pm$0.21 & & 6.84$\pm$0.09   \\
\multicolumn{2}{l}{EW(H$\beta$)(\AA )} & & & 12.7$\pm$0.2 & & 8.2$\pm$0.2 & & 2.6$\pm$0.1& & 9.6$\pm$0.2  \\

\hline
\multicolumn{11}{l}{$^a$in units of 10$^{-14}$ erg s$^{-1}$ cm$^{-2}$}
\end{tabular}
\label{intensities1}
\end{table}
\endlandscape
\newpage
\twocolumn

%%%%%%%%%%%%%%%%%%%%%%%%%%%%%%%%%%%%%%%%%%%%%%%%%%%%%%%%%%%%%%
%
%   TABLA 6 : REDDENING CORRECTED LINE INTENSITIES IN THE OBSERVED CNSFR OF 3351
%
%%%%%%%%%%%%%%%%%%%%%%%%%%%%%%%%%%%%%%%%%%%%%%%%%%%%%%%%%%%%%%

\onecolumn
\landscape

\begin{table}
%\centering
\vspace{3cm}

\caption[]{Reddening corrected emission line intensities for the CNSFR in NGC~3351}
\begin{tabular}{l c c c c c c c c c c}
\hline
Region & & &\multicolumn{2}{c}{R1} & \multicolumn{2}{c}{R2} & \multicolumn{2}{c}{R3} & \multicolumn{2}{c}{R4}\\
            & &  &                &                  &            &                  &                  &                &              &    \\
 $\lambda$ (\AA ) & Line   & f$_\lambda$& F$_\lambda$& I$_\lambda$ &F$_\lambda$& I$_\lambda$&F$_\lambda$& I$_\lambda$&F$_\lambda$& I$_\lambda$ \\
            & &  &                &                  &            &                  &                  &                &              &     \\
 
3727 & [OII]      & 0.271   & 282 $\pm$ 28 & 455 $\pm$ 46 & 172 $\pm$ 17 & 230 $\pm$ 23 & 226 $\pm$ 23 & 336 $\pm$ 34 & 188 $\pm$ 19 & 271 $\pm$ 28 \\ 
4102 & H$_\delta$ & 0.188 &  164$\pm$7  &  229$\pm$9  &  270$\pm$7  &  330$\pm$9  &  255$\pm$6  &  335$\pm$8  &  218$\pm$13 &   281$\pm$16  \\
4340 & H$_\gamma$ & 0.142 &  372$\pm$8  &  478$\pm$10 &  388$\pm$8  &  452$\pm$9  &  434$\pm$7  &  534$\pm$9  &  386$\pm$14 &   467$\pm$17  \\
4686 & HeII       & 0.045 &     --      &     --      &	    --      &	  --      &     --      &     --      &     --      &      --       \\
4861 & H$_\beta$  & 0.000 & 1000$\pm$10 & 1000$\pm$10 & 1000$\pm$10 & 1000$\pm$10 & 1000$\pm$9  & 1000$\pm$9  & 1000$\pm$18 &  1000$\pm$18  \\ 
4959 & [OIII]     &-0.024 &   36$\pm$2  &   34$\pm$2  &   47$\pm$4  &   46$\pm$4  &   27$\pm$2  &   26$\pm$2  &   28$\pm$3  &    27$\pm$3   \\ 
5007 & [OIII]     &-0.035 &  106$\pm$6  &  100$\pm$6  &  140$\pm$6  &  135$\pm$6  &   84$\pm$4  &   79$\pm$4  &   83$\pm$8 &    79$\pm$8  \\
5199 & [NI]       &-0.078 &   49$\pm$13 &   42$\pm$11 &   61$\pm$16 &   56$\pm$14 &   38$\pm$8  &   34$\pm$7  &   61$\pm$17 &    55$\pm$15  \\
5876 & HeI        &-0.209 &   65$\pm$12 &   45$\pm$8  &   63$\pm$27 &   51$\pm$21 &   69$\pm$10 &   51$\pm$7  &   55$\pm$10 &    41$\pm$7   \\
6300 & [OI]       &-0.276 &   24$\pm$6  &   15$\pm$4  &   31$\pm$10 &   23$\pm$7  &   19$\pm$4  &   13$\pm$3  &   19$\pm$5  &    13$\pm$3   \\
6312 & [SIII]     &-0.278 &     --      &     --      &	    --      &	  --      &     --      &     --      &     --      &      --       \\
6548 & [NII]      &-0.311 &  565$\pm$17 &  326$\pm$12 &  476$\pm$16 &  341$\pm$14 &  584$\pm$13 &  372$\pm$11 &  457$\pm$24 &   301$\pm$18  \\
6563 & H$_\alpha$ &-0.313 & 4952$\pm$48 & 2850$\pm$28 & 3981$\pm$42 & 2850$\pm$30 & 4494$\pm$37 & 2850$\pm$24 & 4342$\pm$45 &  2850$\pm$30  \\
6584 & [NII]      &-0.316 & 1739$\pm$30 &  996$\pm$27 & 1489$\pm$21 & 1063$\pm$27 & 1719$\pm$19 & 1086$\pm$23 & 1406$\pm$34 &   919$\pm$35  \\
6678 & HeI        &-0.329 &   16$\pm$4  &    9$\pm$2  &    5$\pm$2  &    3$\pm$2  &   15$\pm$4  &    9$\pm$3  &     --      &      --       \\
6717 & [SII]      &-0.334 &  385$\pm$9  &  214$\pm$7  &  353$\pm$13 &  247$\pm$11 &  347$\pm$6  &  213$\pm$5  &  366$\pm$18 &   233$\pm$14  \\
6731 & [SII]      &-0.336 &  353$\pm$9  &  195$\pm$7  &   337$\pm$12 &   235$\pm$10  &  330$\pm$8  &  203$\pm$6  &  322$\pm$15 &   205$\pm$11  \\
8863 & P11        &-0.546 &   36$\pm$5  &   14$\pm$2  &	    --      &	  --      &     --      &     --      &     --      &      --       \\
9016 & P10        &-0.557 &   53$\pm$5  &   20$\pm$2  &   18$\pm$1  &   10$\pm$1  &   40$\pm$7  &   18$\pm$3  &   24$\pm$4  &    12$\pm$2   \\
9069 & [SIII]     &-0.561 &  264$\pm$15 &   98$\pm$7  &  118$\pm$7  &   65$\pm$5  &  233$\pm$13 &  103$\pm$6  &  139$\pm$13 &    66$\pm$7   \\
9230 & P9         &-0.572 &   92$\pm$6  &   33$\pm$3  &   43$\pm$5  &   23$\pm$3  &   66$\pm$7  &   29$\pm$3  &   66$\pm$51 &    31$\pm$24  \\
9532 & [SIII]     &-0.592 &  697$\pm$30 &  245$\pm$14 &  301$\pm$16 &  160$\pm$11 &  581$\pm$30 &  245$\pm$15 &  346$\pm$21 &   156$\pm$13  \\
9547 & P8         &-0.593 &   89$\pm$5  &   31$\pm$2  &  247$\pm$2  &  131$\pm$5  &  103$\pm$1  &   43$\pm$1  &   92$\pm$7  &    42$\pm$4   \\
          
\hline

c(H$\beta$)   & & & 0.77$\pm$0.03 & & 0.46$\pm$0.03 & & 0.63$\pm$0.02 & & 0.58$\pm$0.04 &  \\
I(H$\beta$)$^a$    & & & & 9.63$\pm$0.10 & & 8.87$\pm$0.09 & & 11.4$\pm$0.11 & & 3.41$\pm$0.06   \\
\multicolumn{2}{l}{EW(H$\beta$)(\AA )} & & & 14.6$\pm$0.2 & & 9.5$\pm$0.1 & & 18.1$\pm$0.3 & & 12.3$\pm$0.3  \\

\hline
\multicolumn{11}{l}{$^a$in units of 10$^{-14}$ erg s$^{-1}$ cm$^{-2}$}
\end{tabular}
\label{intensities2}
\end{table}
\endlandscape
\newpage
\twocolumn

%%%%%%%%%%%%%%%%%%%%%%%%%%%%%%%%%%%%%%%%%%%%%%%%%%%%%%%%%%%%%%
%
%   TABLA 6cont : REDDENING CORRECTED LINE INTENSITIES IN THE OBSERVED CNSFR OF 3351
%
%%%%%%%%%%%%%%%%%%%%%%%%%%%%%%%%%%%%%%%%%%%%%%%%%%%%%%%%%%%%%%

\onecolumn
\landscape

\begin{table}
%\centering
\vspace{3cm}

\contcaption{Reddening corrected emission line intensities for the CNSFR in NGC~3351}
\begin{tabular}{l c c c c c c c c}
\hline
Region & & &\multicolumn{2}{c}{R5} & \multicolumn{2}{c}{R6} & \multicolumn{2}{c}{R7} \\
            & &  &                &                  &            &                  &                  &     \\
 $\lambda$ (\AA ) & Line   & f$_\lambda$& F$_\lambda$& I$_\lambda$ &F$_\lambda$& I$_\lambda$&F$_\lambda$& I$_\lambda$ \\
            & &  &                &                  &            &                  &                  &               \\
 
3727 & [OII]      & 0.271 & 348 $\pm$ 35 & 521 $\pm$ 61 & 350:  & 495:  & 196:  & 284:   \\ 
4102 & H$_\delta$ & 0.188 &     --       &     --       &   85$\pm$6  &  109$\pm$7  &  193$\pm$9  &  249$\pm$11    \\
4340 & H$_\gamma$ & 0.142 &     --       &     --       &  262$\pm$11 &  315$\pm$13 &  400$\pm$10 &  486$\pm$12    \\
4686 & HeII       & 0.045 &     --       &     --       &     --      &     --      &   12$\pm$4  &   12$\pm$5     \\
4861 & H$_\beta$  & 0.000 & 1000$\pm$37  & 1000$\pm$37  & 1000$\pm$20 & 1000$\pm$20 & 1000$\pm$13 & 1000$\pm$13    \\ 
4959 & [OIII]     &-0.024 &  138$\pm$13  &  134$\pm$12  &   32:  &   31:  &   34$\pm$2 &   33$\pm$2    \\ 
5007 & [OIII]     &-0.035 &  409$\pm$37  &  388$\pm$35  &   96: &   91: &  100$\pm$7  &   96$\pm$7     \\
5199 & [NI]       &-0.078 &   78$\pm$33  &   70$\pm$29  &   61$\pm$19 &   55$\pm$17 &   61$\pm$13 &   55$\pm$12    \\
5876 & HeI        &-0.209 &     --       &     --       &     --      &     --      &   36$\pm$8  &   27$\pm$6     \\
6300 & [OI]       &-0.276 &   47$\pm$14  &   31$\pm$10  &     --      &     --      &   14$\pm$5  &    9$\pm$4     \\
6312 & [SIII]     &-0.278 &     --       &     --       &     --      &     --      &     --      &     --         \\
6548 & [NII]      &-0.311 &  622$\pm$43  &  391$\pm$39  &  420$\pm$28 &  282$\pm$20 &  412$\pm$17 &  269$\pm$13    \\
6563 & H$_\alpha$ &-0.313 & 4543$\pm$52  & 2850$\pm$95  & 4259$\pm$44 & 2850$\pm$30 & 4382$\pm$34 & 2850$\pm$22    \\
6584 & [NII]      &-0.316 & 1928$\pm$109 & 1204$\pm$109 & 1272$\pm$55 &  848$\pm$45 & 1289$\pm$29 &  835$\pm$26    \\
6678 & HeI        &-0.329 &     --       &     --       &     --      &     --      &    7$\pm$3  &    4$\pm$2     \\
6717 & [SII]      &-0.334 &  548$\pm$37  &  333$\pm$34  &  275$\pm$18 &  180$\pm$13 &  345$\pm$12 &  218$\pm$9     \\
6731 & [SII]      &-0.336 &  487$\pm$33  &  295$\pm$30  &  248$\pm$17 &  161$\pm$12 &  325$\pm$11 &  205$\pm$8     \\
8863 & P11        &-0.546 &     --       &     --       &     --      &     --      &     --      &     --         \\
9016 & P10        &-0.557 &     --       &     --       &     --      &     --      &   22$\pm$3  &   10$\pm$1     \\
9069 & [SIII]     &-0.561 &  191$\pm$20  &   83$\pm$13  &   89$\pm$13 &   43$\pm$7  &  112$\pm$13 &   52$\pm$6     \\
9230 & P9         &-0.572 &   88$\pm$29  &   38$\pm$13  &   36$\pm$11 &   17$\pm$5  &   66$\pm$7  &   30$\pm$4     \\
9532 & [SIII]     &-0.592 &  415$\pm$39  &  172$\pm$28  &  202$\pm$12 &   95$\pm$8  &  307$\pm$29 &  136$\pm$14    \\
9547 & P8         &-0.593 &     --       &     --       &   74$\pm$11 &   35$\pm$6  &   87$\pm$1  &   39$\pm$2     \\
          
\hline

c(H$\beta$)   & & & 0.65$\pm$0.10 & & 0.56$\pm$0.043 & & 0.60$\pm$0.03 &   \\
I(H$\beta$)$^a$     & & & & 2.04$\pm$0.07 & & 3.26$\pm$0.07 & & 8.05$\pm$0.10   \\
\multicolumn{2}{l}{EW(H$\beta$)(\AA )} & & &  3.1$\pm$0.1 & & 4.9$\pm$0.1 & &  9.0$\pm$0.2  \\

\hline
\multicolumn{9}{l}{$^a$in units of 10$^{-14}$ erg s$^{-1}$ cm$^{-2}$}
\end{tabular}
%\label{intensities2}
\end{table}
\endlandscape
\newpage
\twocolumn

%%%%%%%%%%%%%%%%%%%%%%%%%%%%%%%%%%%%%%%%%%%%%%%%%%%%%%%%%%%%%%
%
%   TABLA 7 : REDDENING CORRECTED LINE INTENSITIES IN THE OBSERVED CNSFR OF 3504
%
%%%%%%%%%%%%%%%%%%%%%%%%%%%%%%%%%%%%%%%%%%%%%%%%%%%%%%%%%%%%%%

\begin{table}
\centering
%\vspace{3cm}

\caption[]{Reddening corrected emission line intensities for the CNSFR in NGC~3504}
\begin{tabular}{l c c c c}
\hline
Region & & &\multicolumn{2}{c}{R3+R4} \\
            & & \\
 $\lambda$ (\AA ) & Line   & f$_\lambda$& F$_\lambda$& I$_\lambda$ \\
            & &  \\
 
3727 & [OII]      & 0.271 & 953 $\pm$ 95 & 1395 $\pm$ 141  \\ 
4102 & H$_\delta$ & 0.188 &  250$\pm$5  &  326$\pm$7  \\
4340 & H$_\gamma$ & 0.142 &   460$\pm$6  &   561$\pm$8  \\
4686 & HeII       & 0.045 &     --      &     --      \\
4861 & H$_\beta$  & 0.000 & 1000$\pm$8  & 1000$\pm$8  \\ 
4959 & [OIII]     &-0.024 &   63$\pm$4  &   61$\pm$4  \\ 
5007 & [OIII]     &-0.035 &  192$\pm$8  &  183$\pm$7  \\
5199 & [NI]       &-0.078 &   69$\pm$15 &   62$\pm$13 \\
5876 & HeI        &-0.209 &   87$\pm$11 &   65$\pm$8  \\
6300 & [OI]       &-0.276 &   80$\pm$7  &   54$\pm$5  \\
6312 & [SIII]     &-0.278 &     --      &     --      \\
6548 & [NII]      &-0.311 &  818$\pm$17 &  529$\pm$13 \\
6563 & H$_\alpha$ &-0.313 & 4423$\pm$29 & 2850$\pm$19 \\
6584 & [NII]      &-0.316 & 2466$\pm$29 & 1583$\pm$30 \\
6678 & HeI        &-0.329 &   17$\pm$4  &   11$\pm$3  \\
6717 & [SII]      &-0.334 &  652$\pm$13 &  408$\pm$10 \\
6731 & [SII]      &-0.336 &  599$\pm$12 &  374$\pm$9  \\
8863 & P11        &-0.546 &     --      &     --      \\
9016 & P10        &-0.557 &   34$\pm$7  &   15$\pm$3  \\
9069 & [SIII]     &-0.561 &  327$\pm$13 &  149$\pm$7  \\
9230 & P9         &-0.572 &   95$\pm$8  &   43$\pm$4  \\
9532 & [SIII]     &-0.592 &  802$\pm$25 &  350$\pm$15 \\
9547 & P8         &-0.593 &     --      &     --      \\
          
\hline

c(H$\beta$)   & & & 0.61$\pm$0.02 & \\
I(H$\beta$)$^a$     & & & & 20.8$\pm$0.17 \\
\multicolumn{2}{l}{EW(H$\beta$)(\AA )} & & & 15.2$\pm$0.2 \\

\hline
\multicolumn{5}{l}{$^a$in units of 10$^{-14}$ erg s$^{-1}$ cm$^{-2}$}
\end{tabular}
\label{intensities3}
\end{table}

%%%%%%%%%%%%%%%%%%%%%%%%%%%%%%%%%%%%%%%%%%%%%%%%%%%%%%%%%%%%
%                                                                             
%             FIGURA  : te ([SIII]) vs log SO23 calibration                     
%                                                                             
%%%%%%%%%%%%%%%%%%%%%%%%%%%%%%%%%%%%%%%%%%%%%%%%%%%%%%%%%%%

%%%%%%%%%%%%%%%%%%%%%%%%%%%%%%%%%%%%%%%%%%%%%%%%%%%%%%%%%%%%
%                                                                             
%             FIGURA  : te ([SIII]) vs log SO23 calibration                     
%                                                                             
%%%%%%%%%%%%%%%%%%%%%%%%%%%%%%%%%%%%%%%%%%%%%%%%%%%%%%%%%%%

%%%%%%%%%%%%%%%%%%%%%%%%%%%%%%%%%%%%%%%%%%%%%%%%%%%%%%%%%%%%
%                                                                             
%             TABLA: Te[SIII] and ionic sulphur abundances                   
%                                                                             
%%%%%%%%%%%%%%%%%%%%%%%%%%%%%%%%%%%%%%%%%%%%%%%%%%%%%%%%%%%
\begin{table*}
\centering
%\vspace{3cm}

\caption[]{Derived electron densities, electron temperatures and ionic abundances from sulphur lines in our observed CNSFR. }
\begin{tabular}{l c c c c c c c c}
\hline
Galaxy & Region & n$_e$ (cm$^{-3}$) & T$_e$([SIII]) (K) & 12+log (S$^+$/H$^+$) & 12+log (S$^{2+}$/H$^+$) & 12+log[ (S$^+$+S$^{2+}$)/H$^+$] \\
\hline
NGC~2903 & R1+R2 & 280 $\pm$  90 & 5731 $\pm$	188  & 6.83 $\pm$ 0.04 &  6.50 $\pm$ 0.05 & 7.01 $\pm$ 0.05\\	
                 & R3       & 230 $\pm$135 & 5737 $\pm$	210  & 6.91 $\pm$ 0.06 &  6.60 $\pm$ 0.06 & 7.09 $\pm$ 0.06 \\  
                 & R4       & 270 $\pm$120 & 6660 $\pm$        331  & 6.68 $\pm$ 0.06 &  6.21 $\pm$ 0.08 & 6.81 $\pm$ 0.06\\   
                 & R6       & 250 $\pm$  80 & 5413 $\pm$	221  & 6.97 $\pm$ 0.05 &  6.64 $\pm$ 0.06 & 7.14 $\pm$ 0.05\\
NGC~3351 & R1       & 360 $\pm$  80 & 5683 $\pm$	188  & 6.74 $\pm$ 0.04 &  6.75 $\pm$ 0.05 & 7.05 $\pm$ 0.04\\
                 & R2       & 440 $\pm$110 & 5405 $\pm$	210  & 6.90 $\pm$ 0.05 &  6.65 $\pm$ 0.06 & 7.09 $\pm$ 0.05\\   
                 & R3       & 430 $\pm$  70 & 5399 $\pm$	210  & 6.84 $\pm$ 0.04 &  6.83 $\pm$ 0.05 & 7.14 $\pm$ 0.05\\  
                 & R4       & 310 $\pm$120 & 5390 $\pm$	271  & 6.87 $\pm$ 0.06 &  6.64$\pm$ 0.07 & 7.07 $\pm$ 0.07\\  
                 & R5       & 360 $\pm$230 & 6177 $\pm$	286  & 6.78 $\pm$ 0.07 &  6.49 $\pm$ 0.09 & 6.97 $\pm$ 0.08\\  
                 & R6       & 360 $\pm$170 & 6297 $\pm$	295  & 6.49 $\pm$ 0.06 &  6.20 $\pm$ 0.07 & 6.67 $\pm$ 0.07\\  
                 & R7       & 410 $\pm$100 & 5537 $\pm$	220  & 6.80 $\pm$ 0.05 &  6.52 $\pm$ 0.07 & 6.99 $\pm$ 0.06\\  
NGC~3504 & R3+R4 & 370 $\pm$  60 & 6288 $\pm$	170  & 6.86 $\pm$ 0.03 &  6.77 $\pm$ 0.03 & 7.12 $\pm$ 0.03\\
\hline
\end{tabular}
\label{ionic-sulphur}
\end{table*}

Once the sulphur ionic abundances have been derived, we have estimated the corresponding oxygen abundances. In order to do that, we have assumed that sulphur and oxygen electron temperatures follow the relation given by Garnett (1992) and confirmed by more recent data \citep{2006MNRAS.372..293H}, and hence we have derived t$_e$([OIII]) according to the expression:
\[
t_e([OIII]) = 1.205 \cdot t_e([SIII]) -0.205
\] 
We have also assumed that t$_e$ ([OII]) $\simeq$ t$_e$([SIII]). The values of T$_e$([OIII]) and the ionic abundances for oxygen are given in Table \ref{ionic-oxygen}. The same comments regarding errors mentioned above apply. Finally, we have derived the N$^{+}$/O$^{+}$ ratio assuming that t$_e$([OII]) $\simeq$ t$_e$([NII]) $\simeq$ t$_e$([SIII]). These values are also listed in Table \ref{ionic-oxygen}.

%%%%%%%%%%%%%%%%%%%%%%%%%%%%%%%%%%%%%%%%%%%%%%%%%%%%%%%%%%%%
%                                                                             
%    TABLA: Te[SIII] and ionic oxygen and nitrogen abundances                   
%                                                                             
%%%%%%%%%%%%%%%%%%%%%%%%%%%%%%%%%%%%%%%%%%%%%%%%%%%%%%%%%%%

\begin{table*}
\centering
%\vspace{3cm}

\caption[]{Derived T$_e$ and ionic abundances for oxygen and nitrogen in our observed CNSFR. }
\begin{tabular}{l c c c c c c c c}
\hline
Galaxy & Region & T$_e$([OIII]) (K) & 12+log (O$^+$/H$^+$) & 12+log (O$^{2+}$/H$^+$) & 12+log[ (O$^+$+O$^{2+}$)/H$^+$] & log(N$^+$/O$^+$)\\
\hline
NGC~2903 & R1+R2 & 4855 $\pm$   226     &  8.55 $\pm$  0.08     &  8.11 $\pm$  0.08    &  8.69 $\pm$  0.08  & -0.37 $\pm$ 0.07 \\
                 & R3       & 4863 $\pm$   253     &  8.55 $\pm$  0.09     &  8.37 $\pm$  0.09    &  8.77 $\pm$  0.09 & -0.36 $\pm$ 0.08  \\
                 & R4       & 5975 $\pm$   399     &  8.42 $\pm$  0.10	 &  8.04 $\pm$  0.14    &  8.57 $\pm$  0.11 & -0.50 $\pm$ 0.08 \\
                 & R6       & 4473 $\pm$   267     &  8.59 $\pm$  0.10	 &  8.33 $\pm$  0.10    &  8.79 $\pm$  0.10 & -0.38 $\pm$ 0.07 \\
NGC~3351 & R1       & 4798 $\pm$   227     &  8.60 $\pm$  0.08	 &  8.06 $\pm$  0.08    &  8.72 $\pm$  0.08 & -0.49 $\pm$ 0.07 \\
                 & R2       & 4463 $\pm$   253     &  8.47 $\pm$  0.09	 &  8.41 $\pm$  0.10    &  8.74 $\pm$  0.09 & -0.23 $\pm$ 0.07 \\
                 & R3       & 4456 $\pm$   253     &  8.64 $\pm$  0.09	 &  8.17 $\pm$  0.10    &  8.77 $\pm$  0.09 & -0.38 $\pm$ 0.07 \\
                 & R4       & 4445 $\pm$   327     &  8.55 $\pm$  0.11	 &  8.19 $\pm$  0.13    &  8.71 $\pm$  0.12 & -0.37 $\pm$ 0.08 \\
                 & R5       & 5393 $\pm$   344     &  8.42 $\pm$  0.10	 &  8.34 $\pm$  0.10    &  8.68 $\pm$  0.10 & -0.37 $\pm$ 0.10 \\
                 & R6       & 5538 $\pm$   356     &  8.34 $\pm$  0.10	 &  7.62 $\pm$  0.17    &  8.42 $\pm$  0.11 & -0.47 $\pm$ 0.08 \\
                 & R7       & 4622 $\pm$   266     &  8.48 $\pm$  0.09	 &  8.15 $\pm$  0.10    &  8.65 $\pm$  0.10 & -0.40 $\pm$ 0.07 \\
NGC~3504 & R3+R4 & 5527 $\pm$   205     &  8.79 $\pm$  0.07	 &  7.94 $\pm$  0.06    &  8.85 $\pm$  0.07 & -0.65 $\pm$ 0.06 \\
\hline
\end{tabular}
\label{ionic-oxygen}
\end{table*}

\begin{figure*}
\centering
\includegraphics[width=.30\textwidth,angle=0]{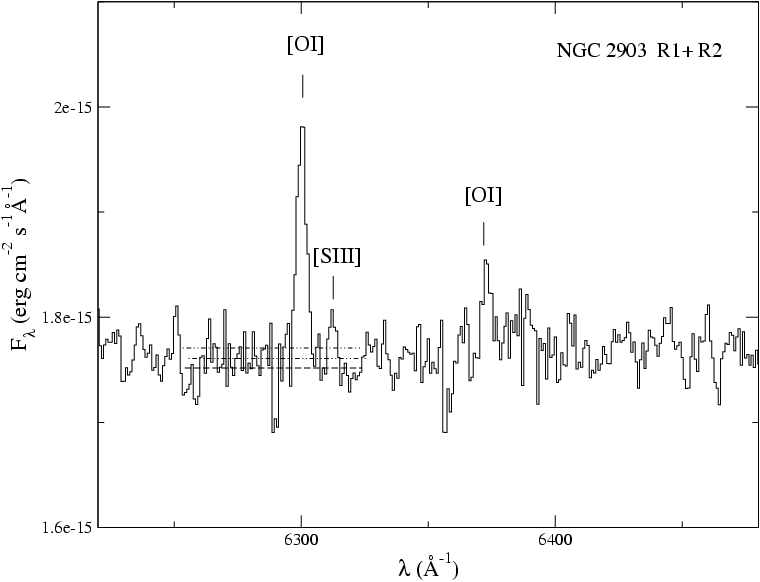}
\caption[]{Spectrum of region R1+R2 in NGC~2903 around [SIII] $\lambda$ 6312 \AA . The way in which the [SIII] line has been measured is shown. The different continuum placements used for the computation of the errors are shown by horizontal lines. }
\label{medidas2}
\end{figure*}

For one of the observed regions, R1+R2 in NGC~2903, the [SIII] $\lambda$ 6312 \AA\ has been detected and measured. 
Although the line intensity ratios for this region presented in Table \ref{intensities1}  have been measured on the combined spectrum of slit positions 1 and 2, the placement of the continuum for the [SIII] 6312 line was very uncertain. Therefore, we have performed the temperature measurement on the spectrum extracted from slit position 1, which shows the best defined continuum. 
On this spectrum we have measured the intensity of the [SIII] $\lambda$ 6312 \AA\ line with respect to H$\alpha$, and those of  [SIII] $\lambda\lambda$ 9069, 9532 \AA\ with respect to P9 $\lambda$ 9329, in order to minimize reddening corrections. The region of this spectrum around the [SIII] $\lambda$ 6312 \AA\ line is shown in Figure \ref{medidas2}  where the horizontal lines show the different placements of the continuum used to measure the line and calculate the corresponding errors.
 The obtained [SIII] nebular to auroral line ratio is: 79.4$\pm$49.1 which gives a T$_e$([SIII])= 8400$^{+ 4650}_{-1250}$K, slightly higher than predicted by the proposed fit. This region is represented as a solid black circle in Figure \ref{te[SIII]cal}.

%%%%%%%%%%%%%%%%%%%%%%%%%%%%%%%%%%%%%%%%%%%%%%%%%%%%%%%%%%%%
%                                                                             
%             FIGURA  : Measurement of the [SIII] 6312  line            
%                                                                             
%%%%%%%%%%%%%%%%%%%%%%%%%%%%%%%%%%%%%%%%%%%%%%%%%%%%%%%%%%%%%%%%%%%%%%%%%%%%%%%

\section{Discussion}

\subsection{Characteristics of the observed CNSFR}
\label{Halpha_fluxes}
Planesas et al. (1997) provide H$\alpha$ fluxes (from images) for all our observed regions. These values are larger than those measured inside our slit by factors between 1.0 and 2.7, depending on the size of the region, something to be expected given the long-slit nature of our observations. We have calculated the H$\alpha$ luminosities for our regions from our observed values, correcting for extinction according to the values found from the spectroscopic analysis. The resulting values are listed in Table \ref{cnsfr_prop}. These values are larger than the typical ones found for disc HII regions and overlap with those measured in HII galaxies. The region with the largest H$\alpha$ luminosity is R3+R4 in NGC~3504, for which a value of 2.02 $\times$ 10$^{40}$ erg s$^{-1}$ is measured. 

We have derived the number of hydrogen ionizing photons from the extinction corrected H$\alpha$ flux as: 
\[ log Q(H^0) = 0.802 \times 10^{49} \left(\frac{F(H\alpha)}{10^{-14}}\right)\left(\frac{D}{10}\right) s^{-1} \]
\noindent where F(H$\alpha$)  is in erg cm$^{-2}$s$^{-1}$ and the distance, D, is in Mpc.

\label{ionization_parameter}
The ionization parameter, u,  can be estimated from the [SII]/[SIII] ratio \citep{1991MNRAS.253..245D} as: 
\[ log u = -1.68 log([SII]/[SIII]) -2.99 \]
and ranges between  -3.12  and -3.98 for our observed CNSFR, in the low side of what is found in disc HII regions, even in the cases of high metallicity \citep[see][]{1991MNRAS.253..245D}.

\label{sizes, filling factors and M(HII)}
From the calculated values of the number of Lyman $\alpha$ photons, Q(H$^0$), ionisation parameter and electron density, it is possible to derive the size of the emitting regions as well as the filling factor \citep[see][]{2002MNRAS.329..315C}. The derived sizes are between 1.5 arcsec for region R3 in NGC~3351 and 5.7 arcsec for region R4 in NGC~2903; these values correspond to linear dimensions between 74 and 234 pc. The derived filling factors are low: between 6 $\times$ 10$^{-4}$ and 1 $\times$ 10$^{-3}$, lower than commonly found in giant HII regions ($\sim$ 0.01). Sizes in arcsec, filling factors and the corresponding masses of ionised hydrogen, M(HII), are given in Table \ref{cnsfr_prop}.

\label{ionising cluster masses}
We have also derived the mass of ionising stars, M*, from the calculated numer of hydrogen ionising photons with the use of evolutionary models of ionising clusters \citep{1994ApJS...91..553G,1996ApJS..107..661S} assuming that the regions are ionisation bound and that no photons are absorbed by dust. A Salpeter IMF with upper and lower mass limits of 100 and 0.8 M$_{\odot}$ respectively, has been assumed. According to these models, a relation exists between the degree of evolution of the cluster, as represented by its H$\beta$ emission line equivalent width and the number of hydrogen ionising photons per unit solar mass \citep{1998Ap&SS.263..143D,2000MNRAS.318..462D}. The ionising cluster masses thus derived are given in Table \ref{cnsfr_prop} and range between  1.1 $\times$ 10$^{5}$ and 4.7 $\times$ 10$^{6}$ M$_{\odot}$. The  measured H$\beta$ equivalent widths, however, are very low and could be reflecting the contribution by underlying non-ionising populations. An alternative way to take into account the cluster evolution  in the derivation of the mass is to make use of the existing relation between the ionisation parameter and the H$\beta$ equivalent width for ionised regions \citep{2006MNRAS.365..454H}. In that case, the derived masses are lower by factors between 1.5 and 15. At any rate, given the assumptions of no dust absorption or photon leakage, these masses represent lower limits.

\subsection{Metallicity estimates}
The abundances we derive using our T$_e$([SIII]) calibration are comparable to those found by Bresolin et al. (2005) for their sample of high metallicity HII regions. Most of our CNSFR show total oxygen abundances, taken to be  O/H= O$^+$/H$^+$ + O$^{2+}$/H$^+$, consistent with solar values within the errors. The region with the highest oxygen abundance is R3+R4 in NGC~3504: 12+log(O/H) = 8.85, about 1.5 solar if the solar oxygen abundance is set at the value derived by Asplund et al. (2005), 12+log(O/H)$_{\odot}$ = 8.66$\pm$0.05. Region R6 in NGC~3351 has the lowest oxygen abundance of the sample, about 0.6 times solar.
In all the observed CNSFR the O/H abundance is dominated by the O$^+$/H$^+$ contribution with 0.18 $\leq$ log(O$^+$/O$^{2+}$) $\leq$ 0.85. This is also the case for high metallicity disc HII regions where these values are even higher. For our observed regions, also the S$^+$/S$^{2+}$ ratios are larger than one, which is at odds with the high metallicity disc HII egions for which, in general, the sulphur abundances are dominated by S$^{2+}$/H$^+$. 

\begin{figure*}
\centering
\vspace*{1.5cm}
\includegraphics[width=.41\textwidth,angle=0]{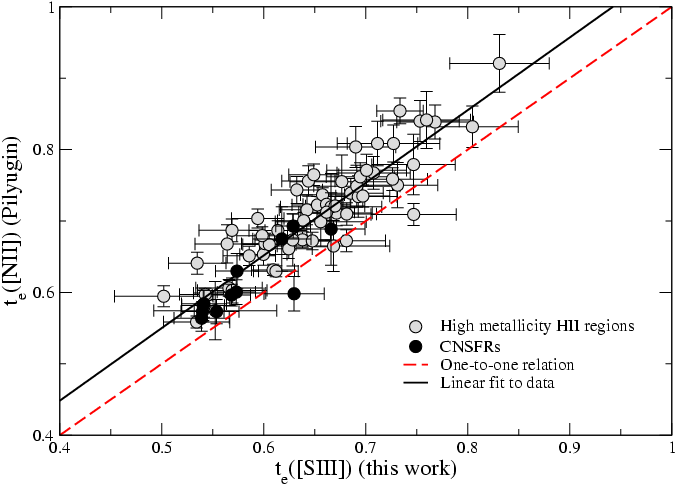}
\caption[]{t$_e$([NII]) derived from Pilyugin's method against the t$_e$([SIII]) derived from the calibration presented in this work. The dashed line shows the one-to-one correspondence.}
\label{Pilyugin}
\end{figure*}

The fact that both O/H and S/H seem to be dominated by the lower ionisation species, can raise concern about our method of abundance derivation, based on the calibration of the T$_e$([SIII]). In a recent article \cite{2007MNRAS.375..685P} addresses this particular problem. He proposes a calibration of the ratio of the [NII] nebular-to-auroral line intensities in terms of those of the nebular oxygen lines. Then the [NII] electron temperature, thought to properly characterise the low ionisation zone of the nebula, can be obtained. We have applied Pilyugin (2007) to our observed CNSFR as well as the high metallicity HII sample. Figure \ref{Pilyugin} shows the derived t$_e$([NII]) following Pilyugin's method against the t$_e$([SIII]) derived from our calibration. The (red) dashed line shows the one-to-one relation while the (black) solid line shows the actual fit to all the data. It can be seen that, in what regards CNSFR, both temperatures are very similar, with the [NII] temperature being, in average, 500 K higher than that of [SIII].
% although the CNSFR lie closer to the one-to-one relation. 
This difference is of the same size of the average errors in the measured temperatures entering the calibrations, therefore we can assume that our derived t$_e$([SIII]) characterises the low ionisation zone at least as well as the t$_e$([NII]) derived applying Pilyugin's method.
 
Concerning relative abundances, with our analysis it is possible to derive the relative N/O value, assumed to be equal to the N$^{+}$/O$^+$ ratio. These values are in all cases larger than the solar one (log(N/O)$_{\odot}$= -0.88; Asplund et al. 2005) by factors between 1.7 (R3+R4 in NGC~3504) and 4.5 (R2 in NGC~3351) which are amongst the highest observed N/O ratios \citep[see e.g.][]{2006MNRAS.372.1069M}.
Regarding S, if -- given the low excitation of the observed regions -- the fraction of S$^{3+}$ is assumed to be negligible, the S/O ratio can be obtained as:
\[ \frac{S}{O} = \frac{S^{+}+S^{2+}}{O^{+}+O^{2+}} \]
The values of log(S/O) span a very narrow range between -1.76 and -1.63, that is between 0.6 and 0.8 of the solar value \citep[log(S/O)$_{\odot}$ = -1.52;][]{2005ASPC..336...25A}.

%%%%%%%%%%%%%%%%%%%%%%%%%%%%%%%%%%%%%%%%%%%%%%%%%%%%%%%%%%%%
%                                                                             
%             FIGURA  : te([NII]) vs te([SIII])                
%                                                                             
%%%%%%%%%%%%%%%%%%%%%%%%%%%%%%%%%%%%%%%%%%%%%%%%%%%%%%%%%%%

%%%%%%%%%%%%%%%%%%%%%%%%%%%%%%%%%%%%%%%%%%%%%%%%%%%%%%%%%%%

\begin{table*}
\centering
%\vspace{3cm}

\caption[]{General properties of the observed CNSFR. }
\begin{tabular}{l c c c c c c c c c}
\hline
Galaxy& Region   &  F(H$\alpha$) & L(H$\alpha$) & Q(H$^0$)	& log u &  Diameter & $\epsilon$ & M$*$ &   M(HII) \\
        &  & (erg cm$^{-2}$ s$^{-1}$) & (erg s$^{-1}$) &  (photons s$^{-1}$) &  &  (arcsec) &   &  (M$_{\odot}$) & (M$_{\odot}$) \\
\hline

NGC~2903 & R1+R2 & 3.06E-13 & 2.71E+39 &	1.98E+51 & -3.69  &     4.66	& 5.93E-04  &  7.39E+05 &	1.44E+04\\
                 & R3       & 3.16E-14 & 2.80E+38 &	2.05E+50 & -3.65  &     1.59	& 2.29E-03  &  1.11E+05 &	1.81E+03\\
                 & R4       & 2.23E-13 & 1.98E+39 &	1.45E+51 & -3.98  &     5.70	& 2.54E-04  &  2.11E+06 &	1.09E+04\\
                 & R6       & 1.48E-13 & 1.31E+39 &	9.60E+50 & -3.66  &     3.33	& 9.89E-04  &  4.55E+05 &	7.79E+03\\
NGC~3351 & R1       & 2.19E-13 & 2.65E+39 &	1.94E+51 & -3.12  &     1.81	& 3.76E-03  &  6.40E+05 &	1.09E+04\\
                 & R2       & 1.61E-13 & 1.95E+39 &	1.42E+51 & -3.55  &     2.29	& 9.08E-04  &  6.80E+05 &	6.56E+03\\
                 & R3       & 2.35E-13 & 2.84E+39 &	2.08E+51 & -3.12  &     1.72	& 3.31E-03  &  5.70E+05 &	9.80E+03\\
                 & R4       & 6.84E-14 & 8.29E+38 &	6.06E+50 & -3.49  &     1.66	& 2.04E-03  &  2.32E+05 &	3.96E+03\\
                 & R5       & 4.25E-14 & 5.15E+38 &	3.76E+50 & -3.65  &     1.47	& 1.37E-03  &  4.71E+05 &	2.12E+03\\
                 & R6       & 6.43E-14 & 7.79E+38 &	5.70E+50 & -3.65  &     1.81	& 1.11E-03  &  4.81E+05 &	3.21E+03\\
                 & R7       & 1.60E-13 & 1.94E+39 &	1.42E+51 & -3.58  &     2.47	& 8.31E-04  &  7.11E+05 &	7.03E+03\\
NGC~3504 & R3+R4 & 4.20E-13 & 2.02E+40 &	1.47E+52 & -3.32  &     3.11	& 6.76E-04  &  4.70E+06 &	8.07E+04\\
\hline
\end{tabular}
\label{cnsfr_prop}
\end{table*}  

%%%%%%%%%%%%%%%%%%%%%%%%%%%%%%%%%%%%%%%%%%%%%%%%%%%%%%%%%%%%
%                                                                             
%             FIGURA  : histograms of O23 and N2                    
%                                                                             
%%%%%%%%%%%%%%%%%%%%%%%%%%%%%%%%%%%%%%%%%%%%%%%%%%%%%%%%%%%
\begin{figure*}
\centering
\vspace{0.5cm}
\includegraphics[width=.41\textwidth,angle=0]{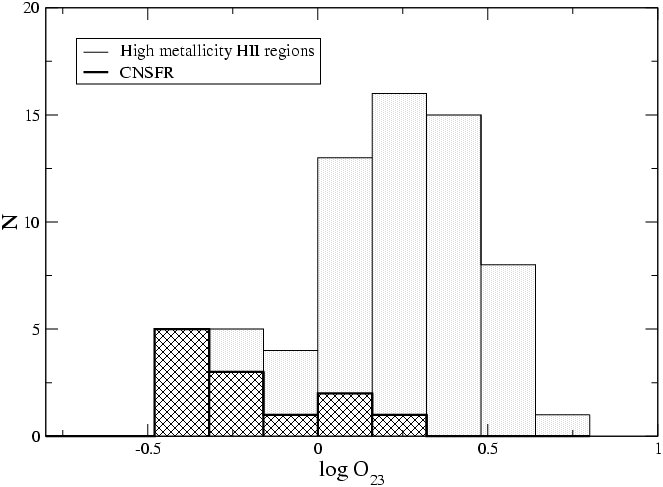}
\hspace{0.3cm}
\includegraphics[width=.41\textwidth,angle=0]{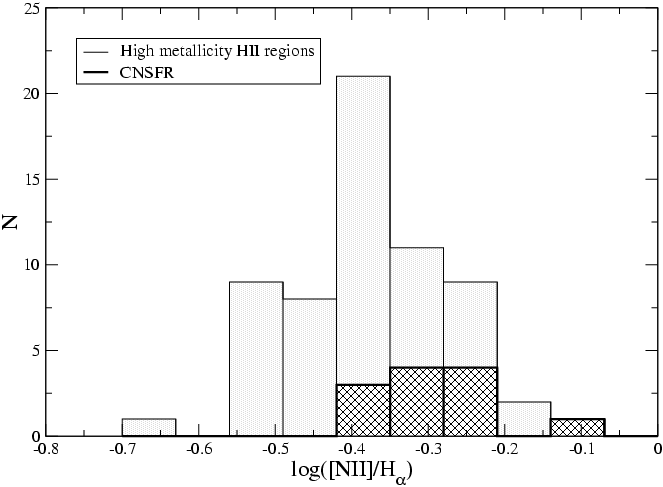}
\caption[]{Distribution of the empirical abundance parameters O$_{23}$ (left) and N2 (right) for the observed CNSFR and the sample of high metallicity disc HII regions.}
\label{histo_O23_N2}
\end{figure*}

\subsection{Comparison with high metallicity HII regions}
\begin{figure*}
\centering
\vspace{0.5cm}
\includegraphics[width=.41\textwidth,angle=0]{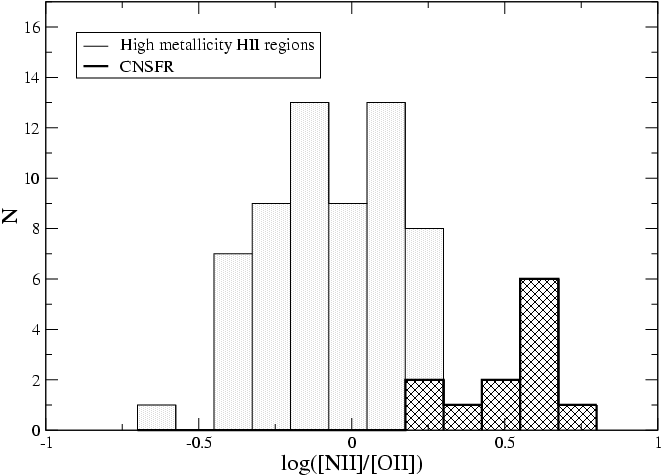}
\caption[]{Distribution of the [NII]/[OII] ratio for the observed CNSFR (dark) and the sample of high metallicity disc HII regions (light).}
\label{N2-O2}
\end{figure*}

The observed CNSFR being of high metallicity show however marked differences with respect to high metallicity disc HII regions. Even though their derived oxygen and sulphur abundances are similar, they show values of the O$_{23}$ and the N2 parameters whose distributions are shifted to lower and higher values respectively with respect to the high metallicity disc sample (Figure \ref{histo_O23_N2}). Hence, if pure empirical methods were used to estimate the oxygen abundances for these regions, higher values would in principle be obtained. This would seem to be in agreement with the fact that CNSFR, when compared to the disc high metallicity regions, show the highest [NII]/[OII] ratios.  Figure \ref{N2-O2} shows indeed a bi-modal distribution of the [NII]/[OII] ratio in disc and circumnuclear HII regions.

%%%%%%%%%%%%%%%%%%%%%%%%%%%%%%%%%%%%%%%%%%%%%%%%%%%%%%%%%%%%
%                                                                             
%             FIGURA  : histogram of N2/O2                   
%                                                                             
%%%%%%%%%%%%%%%%%%%%%%%%%%%%%%%%%%%%%%%%%%%%%%%%%%%%%%%%%%%

\begin{figure*}
\centering
\vspace{0.5cm}
\includegraphics[width=.41\textwidth,angle=0]{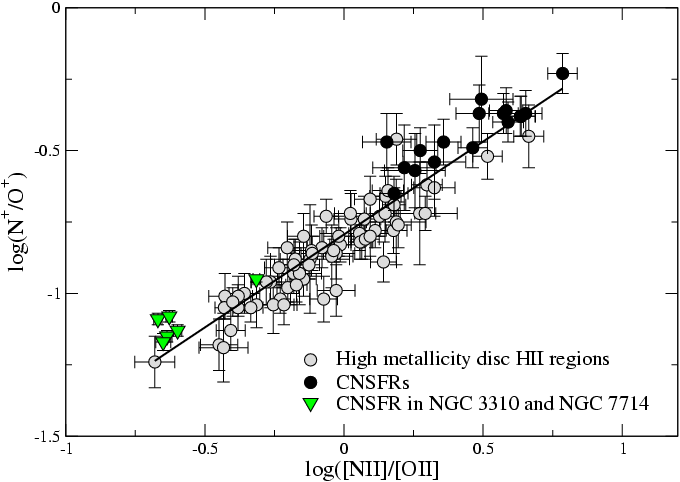}
\caption[]{The relation between the [NII]/[OII] emission line intensity ratio and the N$^+$/O$^+$ ratio for CNSFR (black  cicles) and high metallicity HII regions (grey circles). Included also (downward triangles) are lower metallicity CNSFR in NGC~3310 \citep{1993MNRAS.260..177P} and NGC~7714 \citep{1995ApJ...439..604G}.}
\label{NoverO}
\end{figure*}

A good correlation has been found to exist between the [NII]/[OII] ratio and the N$^+$/O$^+$ ionic abundance ratio, which in turn can be assumed to trace the N/O ratio  \citep{2005MNRAS.361.1063P}. This relation is shown in Figure \ref{NoverO} for the observed CNSFR (black circles) and the high metallicity HII region sample (grey circles). The CNSFR sample has been enlarged with two circumnuclear regions of NGC~1097 observed by \cite{1984MNRAS.210..701P} and other three observed in NGC~5953 by \cite{1996MNRAS.281..781G}, all of them of high metallicity.  Also shown are data of some CNSFR in two peculiar galaxies: NGC~3310 \citep{1993MNRAS.260..177P} and NGC~7714 \citep{1995ApJ...439..604G} of reported lower metallicity. In all the cases, ionic and total abundances have been derived following the same methods as in the CNSFR in the present study and described in section 5. In the case of the latter regions, abundances derived in this way are larger than derived from direct determinations of t$_e$([OIII]) since our estimated values for this temperature are systematically lower by about 1200 K. This could be due to the fact that our semi-empirical calibration has been actually derived for high metallicity regions. However, the values of log(SO$_{23}$) for these regions are between -0.44 and -0.85 and therefore within the validity range of the calibration and we have preferred to use the same method for consistency reasons.

%%%%%%%%%%%%%%%%%%%%%%%%%%%%%%%%%%%%%%%%%%%%%%%%%%%%%%%%%%%%
%                                                                             
%             FIGURA  : N/O vs [NII]/[OII]                 
%                                                                             
%%%%%%%%%%%%%%%%%%%%%%%%%%%%%%%%%%%%%%%%%%%%%%%%%%%%%%%%%%%

\begin{figure*}
\centering
\vspace{0.5cm}
\includegraphics[width=.41\textwidth,angle=0]{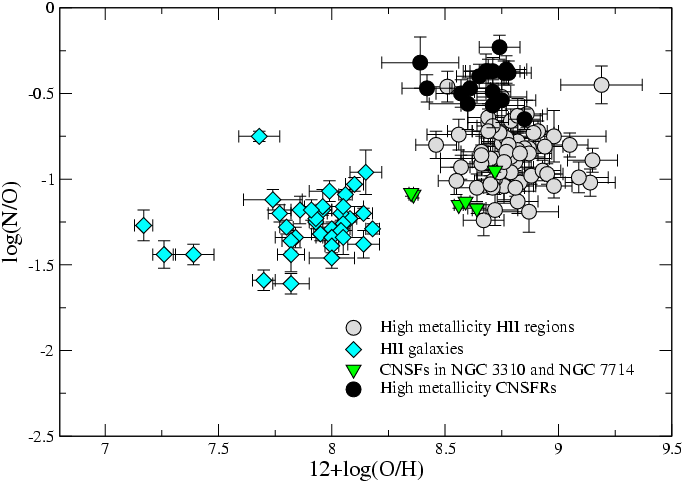}
\caption[]{Relation between the N/O ratio and the O/H abundance for CNSFR (black circles and green upside down triangles for high and low metallicities respectively), high metallicity HII regions (grey circles) and HII galaxies (cyan  diamonds).}
\label{NoverO-O}
\end{figure*}

We can see that a very tight correlation exists which allows to estimate the N/O ratio from the measured [NII] and [OII] emission line intensities. In this relation, high metallicity regions and CNSFR seem to follow a sequence of increasing N/O ratio. A linear regression fit to the data yields the expression:
\[ log(N/O) = (0.65 \pm 0.02) log ([NII]/[OII]) - (0.79 \pm 0.01) \]
This relation is shallower than that found in \cite{2005MNRAS.361.1063P} for a sample that did not include high metallicity HII regions. The N/O ratio for our observed CNSFR is shown in Figure \ref{NoverO-O} against their oxygen abundance together with similar data for the high metallicity HII region and HII galaxy samples. It can be seen that all the CNSFR show similar oxygen abundances, with the mean value being lower than that shown by high metallicity disc HII regions, but the observed CNSFR show larger N/O ratios and they do not seem to follow the trend of N/O vs O/H which marks the secondary behaviour of nitrogen.

%%%%%%%%%%%%%%%%%%%%%%%%%%%%%%%%%%%%%%%%%%%%%%%%%%%%%%%%%%%%
%                                                                             
%             FIGURA  : N/O vs O/H             
%                                                                             
%%%%%%%%%%%%%%%%%%%%%%%%%%%%%%%%%%%%%%%%%%%%%%%%%%%%%%%%%%%

\label{What's the true metalicity of CNSFR?}
The values of the oxygen abundance that we find for the CNSFR in NGC~3351 are equal within the errors, with a mean value 12+log(O/H) = 8.70 $\pm$ 0.10, except for region R6 which shows an O/H abundance lower by a factor of about 2. It is worth noting that this is the region with the highest O$^+$/O ratio: 0.83 and a very low O$^{2+}$/H$^+$ value: 12+log(O$^{2+}$/H$^+$) = 7.62 $\pm$ 0.17. The average value for the rest of the regions is in agreement with the central abundance found by \cite{2006MNRAS.367.1139P} by extrapolating the galaxy oxygen abundance gradient, 12+log(O/H) = 8.74 $\pm$ 0.02. This value is somewhat lower than that previously estimated by \cite{2004A&A...425..849P}: 8.90 from the same data but following a different method of analysis. In that same work, the quoted central abundance for NGC~2903 is 12+log(O/H) = 8.94. Our results for regions R1+R2, R3 and  R6 in this galaxy yield an average value of 8.75 $\pm$ 0.09 lower than theirs by 0.2 dex. Region R4 shows a lower oxygen abundance but still consistent within the errors with the average. 
Values of N/O ratios for the centres of NGC~3351 and NGC~2903 similar to those found here (-0.33 and -0.35 respectively) are quoted by \cite{2004A&A...425..849P}. 
%Interestingly, this region shows, at the same time,  the lowest N/O ratio. 

\begin{figure*}
\centering
\vspace{0.5cm}
\includegraphics[width=.41\textwidth,angle=0]{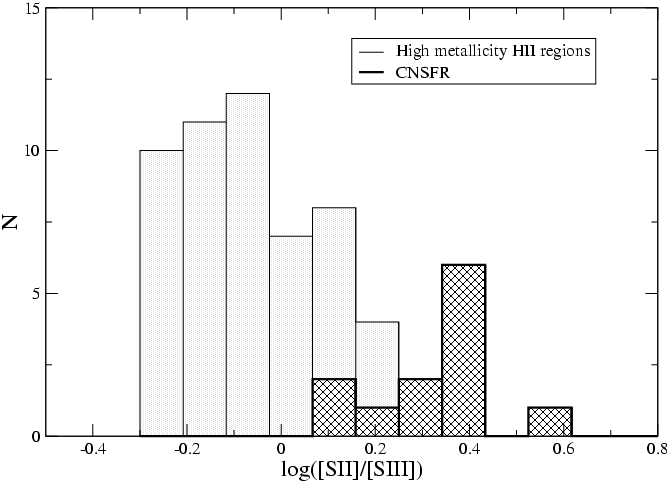}
\hspace{0.2cm}
\includegraphics[width=.41\textwidth,angle=0]{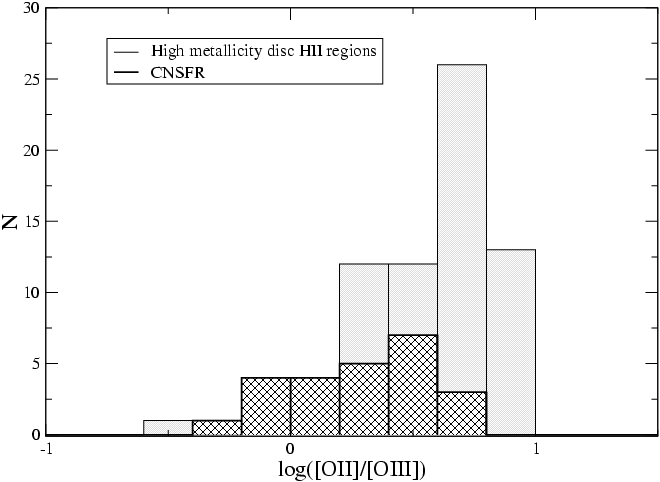}
\caption[]{Distribution of the [SII]/[SIII] (left) and [OII]/[OIII] (right) ratios for the observed CNSFR (dark) and the sample of high metallicity disc HII regions (light).}
\label{S2-S3}
\end{figure*}

\label{ionisation parameter}
Another difference between the high metallicity circumnuclear and disc regions is related to their average ionisation parameter. The left panel of Figure \ref{S2-S3} shows the distribution of the [SII]/[SIII] ratio for the two samples. The [SII]/[SIII] ratio has been shown to be a good ionisation parameter indicator for moderate to high metallicities \citep{1991MNRAS.253..245D} with very little dependence on metallicity or ionisation temperature. It can be seen that all the CNSFR observed show large [SII]/[SIII] ratios which imply extremely low ionisation parameters. On the other hand, a different answer would be found if the [OII]/[OIII] parameter, also commonly used as ionisation parameter indicator, was used. In this case, CNSFR and high metallicity HII regions show a much more similar distribution, with CNSFR showing slightly lower values of [OII]/[OIII] (Figure \ref{S2-S3}, right panel). 

It should be noted the dependence of the [OII]/[OIII] parameter on metallicity due to the presence of opacity edges of various abundant elements (O$^+$, Ne$^+$, C$^{2+}$, N$^{2+}$) in the stellar atmospheres  that can combine to substantially modify the stellar flux of high abundance stars at energies higher than 35-30 eV and then produce a lower [OIII] emission \citep{1976ApJ...208..336B}. However, if the CNSFR were ionised by stars of a higher metallicity than those in disc HII regions this effect would go in the direction of producing higher [OII]/[OIII] ratios for the CNSFR, and ionisation parameters derived from  [OII]/[OIII] ratios would be found to be lower than those derived from [SII]/[SIII] ratios, contrary to what is actually observed. 

%%%%%%%%%%%%%%%%%%%%%%%%%%%%%%%%%%%%%%%%%%%%%%%%%%%%%%%%%%%%
%                                                                             
%             FIGURA  : histograms of [SII]/[SIII]    and [OII]/[OIII]               
%                                                                             
%%%%%%%%%%%%%%%%%%%%%%%%%%%%%%%%%%%%%%%%%%%%%%%%%%%%%%%%%%%

\begin{figure*}
\centering
\includegraphics[width=.41\textwidth,angle=0]{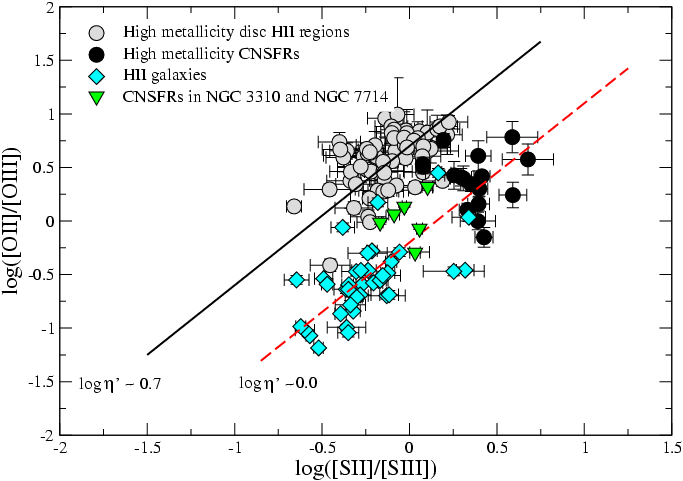}
\caption[]{The $\eta$ ' plot: the [OII]/[OIII] ratio vs [SII]/[SIII] ratio for different ionised regions: high metallicity CNSFR (black circles), low metallicity CNSFR (green upside down triangles), high metallicity disc HII regions (grey circles and HII galaxies (cyan diamonds). }
\label{eta-prime-plot}
\end{figure*}

\label{Ionisation structure}
The ionisation structure can provide important information about the characteristics of the ionising source. A diagram of the emission line ratios [OII]/[OIII] vs [SII]/[SIII], in particular, works as a diagnostics for the nature and temperature of the radiation field. This diagram was used in \cite{1985MNRAS.212..737D} in order to investigate the possible contributions by shocks in CNSFR and LINERs and is the basis of the definition of the $\eta$' parameter \citep{1988MNRAS.231..257V}. The $\eta$' parameter, defined as: 
\[ \eta ' = \frac{[OII]\lambda\lambda 3727,29 / [OIII]\lambda\lambda 4959,5007}{[SII]\lambda\lambda 6716,6731 /[SIII] \lambda\lambda 9069,9532} \] 
is a measure of the ``softness" of the ionising radiation  and increases with decreasing ionising temperature. 
The ``$\eta$' plot" is shown in Figure \ref{eta-prime-plot}. In this plot, diagonal lines of slope unity would show the locus of ionised regions with constant values of $\eta$'. The lines shown in the plot have slope 1.3 reflecting the second order dependence of $\eta$' on ionisation parameter \citep{1991MNRAS.253..245D}. In this graph CNSFR are seen to segregate from disc HII regions. The former cluster around the value of log$\eta$' = 0.0 (T${ion}\sim$ 40,000 K) while the latter cluster around log $\eta$' = 0.7 (T${ion}\sim$ 35,000 K). Also shown are the data corresponding to HII galaxies. Indeed, CNSFR seem to share more
the locus of the HII galaxies than that of disc HII regions.

%%%%%%%%%%%%%%%%%%%%%%%%%%%%%%%%%%%%%%%%%%%%%%%%%%%%%%%%%%%
%
%                                    Figure eta prime
%
%%%%%%%%%%%%%%%%%%%%%%%%%%%%%%%%%%%%%%%%%%%%%%%%%%%%%%%%%%%

Given the problems involved in he measurement of the [OII] emission lines mentioned in section 4.2 it is interesting to see the effects that a possible underestimate of the intensity of this line would have on our analysis.  The underestimate of the [OII] line leads to an overestimate of the SO$_{23}$ parameter and hence to an underestimate of the [SIII] electron temperature. This lower temperature in turn leads to higher O/H abundances. 
      Therefore, larger values of the intensity of the [OII] line would lead to lower O/H abundances. However, not by a large amount. Increasing the [OII] intensity by a factor of 2, would decrease the SO$_{23}$ parameter by a factor between 1.5 and 1.8 depending on the region, which would lead to increased [SIII] electron temperatures by between 500 and 1000 K.
The corresponding O$^+$/H$^+$ would be only slightly smaller, by about 0.10 dex, as would also be the total O/H abundances which is dominated by O$^+$/H$^+$. The O$^{++}$/H$^+$ ionic ratio however would be lower by between 0.3 and 0.4 dex, which would produce an ionisation structure more similar to what is found in disc HII regions. The sulphur ionic and total abundances would be decresed by about 0.17 dex. Finally, the N+/O+ ratios would be decreased by about 0.17 dex, while the S/O ratios would remain almost unchanged. In the ``$\eta$' plot" (Fig. \ref{eta-prime-plot}) the data points corresponding to our observed CNSFR 
would move upwards by 0.30 dex.
However, we do not find any compelling reason why the [OII] intensities should be larger than measured by such a big amount (see section 4.2 and Fig. \ref{medidas1}).

One possible concern about these CNSFR is that, given their proximity to the galactic nuclei, they could be affected by  hard radiation coming from a low luminosity AGN. NGC~3551 shows a faint UV core . However, the IUE spectrum that covers the whole central starforming ring, shows broad absorption lines of SiIV $\lambda$ 1400 \AA\ and CIV $\lambda$ 1549~\AA\ typical of young stars of high metallicity \citep{1997ApJ...484L..41C}. They are consistent with a total mass of 3 $\times$ 10$^5$ M$_{\odot}$ of recently formed stars (4-5 Myr). This is of the order of our derived values for single CNSFR in this galaxy. Therefore, no signs of activity are found for this nucleus nor are they reported for the other two galaxy nuclei. On the other hand, the HeII $\lambda$ 4686 \AA\ line is measured in regions R1+R2 and R6 of NGC~2903  and in region R7 in NGC~3351. In the first region, there is some evidence for the presence of WR stars (Castellanos et al. 2002). For the other two, that presence is difficult to assess due to the difficulty in placing the contiuum for which a detailed modelling of the stellar population is needed. 

\begin{figure*}
\centering
\includegraphics[width=.41\textwidth,angle=0]{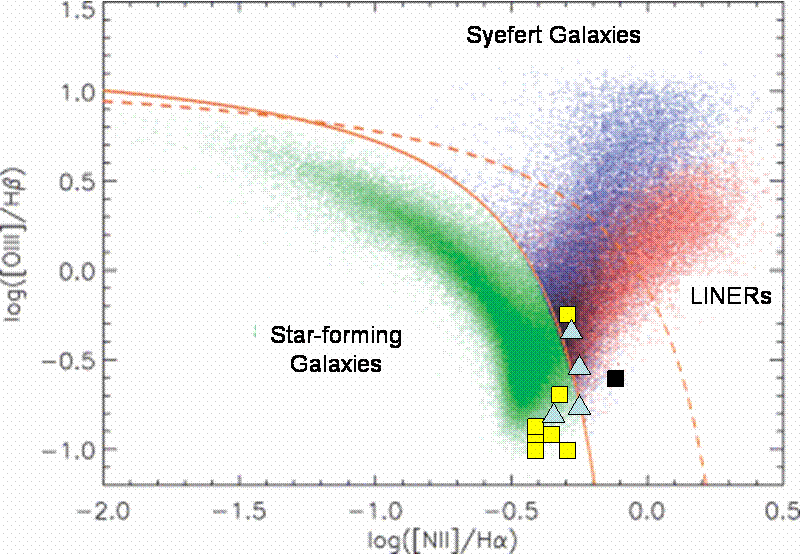}
\caption[]{The [NII]/H$\alpha$  vs [OIII]/H$\beta$ diagnostics for emission line objects in the SDSS (adapted from \cite{2006MNRAS.371.1559G}). The location of our observed CNSFR is shown overplotted. Light (blue) triangles correspond to NGC~2903, light (yellow) squares correspond to NGC~3351 and the black square corresponds to NGC~3504. Solid \citep{2001ApJ...556..121K} and dashed \citep{2003MNRAS.346.1055K} lines
separate the starforming from the active nucleus galaxies regions.
}
\label{diagnostico}
\end{figure*}

 Alternatively, the spectra of  these regions harbouring massive clusters of young stars might be affected by the presence of shocked gas. Diagnostic diagrams of the kind presented by \cite{1981PASP...93....5B} can be used to investigate the possible contribution by either a hidden AGN or by the presence of shocks to the emission line spectra of the observed CNSFR. Figure \ref{diagnostico} shows one of these diagnostic diagrams, log ([NII]/H$\alpha$) vs log ([OIII])/H$\beta$, for our CNSFR: light squares for NGC~3351, triangles for NGC~2903 and dark square for NGC~3504. The figure has been adapted from \cite{2006MNRAS.371.1559G} and shows the location of emission line galaxies in the Sloan Digital Sky Survey (SDSS). Dashed and solid lines correspond to the boundary between Active Galactic Nuclei (AGN) and HII galaxies defined by \cite{2001ApJ...556..121K} and  \cite{2003MNRAS.346.1055K} respectively. Some of our observed CNSFR are found close to the transition zone between HII region and LINER spectra but only R3+R4 in NGC~3504 may show a hint of a slight contamination by shocks. This region, the most luminous in our sample and also the one with the higest abundance,  should be studied in more detail.

A final remark concerns the gas kinematics in CNSFR. In a recent work \cite{2007MNRAS.378..163H} have studied the kinematics of gas and stars in the CNSFR of NGC~3351 finding two different components for the ionised gas in H$\beta$ and [OIII] emission: a ``broad component" with a velocity dispersion similar to that measured for the stars,
and a ``narrow component" with a dispersion lower than the stellar one by about 30\,km\,s$^{-1}$. Obviously the abundance analysis and the location of these regions on diagnostic diagrams would be affected if more than one velocity component in the ionised gas corresponding to kinematically distinct systems are present. 

%%%%%%%%%%%%%%%%%%%%%%%%%%%%%%%%%%%%%%%%%%%%%%%%%%%%%%%%%%%
%
%                                          Figure diagnostics
%
%%%%%%%%%%%%%%%%%%%%%%%%%%%%%%%%%%%%%%%%%%%%%%%%%%%%%%%%%%%

\section{Summary and conclusions}
We have obtained spectro-photometric observations in the optical ($\lambda\lambda$ 3650-7000 \AA ) and far-red ($\lambda\lambda$ 8850-9650 \AA ) wavelength ranges of 12 circumnuclear HII regions in the early type spiral galaxies: NGC~2903, NGC~3351 and NGC~3504. These regions were expected to be amongst the highest metallicity regions as corresponds to their position near the galactic centre. At the same time, this position implies a substantial contribution by the bulge stellar population to their spectra which represents a major observational problem and compromises the reliability of the emission line intensities. Its proper subtraction however requires the detailed modelling of the stellar population and the disentangling of the contribution by the bulge and by any previous stellar generations in the CNSFR themselves. Almost all these regions show the presence of the CaT lines in the far red (see H\"agele et al. 2007) some of them with equivalent widths that suggest a certain contribution by red supergiant stars. A detailed study of the ionising and non ionising stellar populations of this regions will be presented in a forthcoming paper. 
In the case of the Balmer lines, the presence of wide wings in their profiles, allows to perform a two-component -- emission and absorption -- gaussian fit in order to correct the Balmer emission lines for this effect. 

We have derived the characteristics of the observed CNSFR in terms of size, H$\alpha$ luminosities and ionising cluster masses. The derived sizes are between 1.5 arcsec  and 5.7 arcsec which correspond to linear dimensions between 74 and 234 pc. The derived filling factors, between 6 $\times$ 10$^{-4}$ and 1 $\times$ 10$^{-3}$, are lower than commonly found in giant HII regions ($\sim$ 0.01). H$\alpha$ luminosities are larger than the typical ones found for disc HII regions and overlap with those measured in HII galaxies. The region with the largest H$\alpha$ luminosity is R3+R4 in NGC~3504, for which a value of 2.02 $\times$ 10$^{40}$ is measured.  Ionising cluster masses range between  1.1 $\times$ 10$^{5}$ and 4.7 $\times$ 10$^{6}$ M$_{\odot}$ but could be lower by factors between 1.5 and 15 if the contribution by the underlying stellar population is taken into account.

The low excitation of the regions, as evidenced by the weakness of the [OIII] $\lambda$ 5007 \AA\ line, precludes the detection and measurement of the auroral [OIII] $\lambda$ 4363 \AA\ necessary for the derivation of the electron temperature. Only for one of the regions, the [SIII] $\lambda$ 6312  \AA\ line was detected providing, together with the nebular [SIII] lines at $\lambda\lambda$ 9069, 9532 \AA\ , a value of the electron temperature of T$_e$([SIII])= 8400$^{+ 4650}_{-1250}$K. A new method for the derivation of sulphur abundances was developed based on the calibration of the [SIII] electron temperature vs the empirical parameter SO$_{23}$ defined as the quotient of the oxygen and sulphur abundance parameters O$_{23}$ and S$_{23}$ and the further assumption that T([SIII]) $ \simeq $ T([SII]). Then the oxygen abundances and the N/O and S/O ratios can also be derived. 

The derived oxygen abundances are comparable to those found in high metallicity disc HII regions from direct measurements of electron temperatures and consistent with solar values within the errors. The region with the highest oxygen abundance is R3+R4 in NGC~3504, 12+log(O/H) = 8.85, about 1.5 solar if the solar oxygen abundance is set at the value derived by Asplund et al. (2005), 12+log(O/H)$_{\odot}$ = 8.66$\pm$0.05. Region R7 in NGC~3351 has the lowest oxygen abundance of the sample, about 0.6 times solar.
In all the observed CNSFR the O/H abundance is dominated by the O$^+$/H$^+$ contribution, as is also the case for high metallicity disc HII regions. For our observed regions, however also the  S$^+$/S$^{2+}$ ratio is larger than one, different from the case of high metallicity disc HII regions for which, in general, the sulphur abundances are dominated by S$^{2+}$/H$^+$. 
The derived N/O ratios are in average larger than those found in high metallicity disc HII regions and they do not seem to follow the trend of N/O vs O/H which marks the secondary behaviour of nitrogen. On the other hand, the S/O ratios span a very narrow range between 0.6 and 0.8 of the solar value.
 
When compared to high metallicity disc HII regions, CNSFR show values of the O$_{23}$ and the N2 parameters whose distributions are shifted to lower and higher values respectively, hence, even though their derived oxygen and sulphur abundances are similar, higher values would in principle be obtained for the CNSFR  if pure empirical methods were used to estimate abundances. CNSFR also show lower ionisation parameters than their disc counterparts, as derived from the [SII]/[SIII] ratio. Their ionisation structure also seems to be different with CNSFR showing radiation field properties more similar to HII galaxies than to disc high metallicity HII regions. The possible contamination of their spectra from hidden low luminosity AGN and/or shocks, as well as the probable presence  of more than one velocity component in the ionised gas corresponding to kinematically distinct systems, should be further investigated. 

\section*{Acknowledgements}

The WHT is operated in the island of La Palma by the Isaac Newton Group
in the Spanish Observatorio del Roque de los Muchachos of the Instituto
de Astrof\'\i sica de Canarias. We thank the Spanish allocation committee
(CAT) for awarding observing time.

This work has been partially supported by DGICYT grant AYA-2004-02860-C03. GH
acknowledges support from the Spanish MEC through FPU grant AP2003-1821. 
AID acknowledges support from  the Spanish MEC through a
sabbatical grant PR2006-0049. Also, partial support from the Comunidad de 
Madrid under grant S-0505/ESP/000237 (ASTROCAM) is acknowledged. Support from
the Mexican Research Council (CONACYT) through grant 49942 is acknowledged by
ET. We thank the hospitality of the Institute of Astronomy, Cambridge,
where most of this paper was written. 

We would like to thank Mike Beasley for providing the digital spectrum of the M~31 cluster, 
Roberto Cid Fernandes for very helpful discussions concerning subtraction procedures 
of  the underlying absorptions and Roberto Terlevich for a careful reading of this manuscript.
We also thank an anonimous referee for a very careful review of this work which lead to the 
improvement of the paper.

This work is dedicated to the memory of  Bernard Pagel who was always a source of inspiration and stimuli for us and with whom we discussed many of the matters addressed in this paper over the last twenty years or so.

\bibliographystyle{mn2e}
\bibliography{cn_HII_rev}
\end{document}